\documentclass[12pt]{article}

\usepackage{thumbpdf,lmodern}
\usepackage{amsfonts}
\usepackage{url}
\usepackage{Sweave}

\newcommand{\rd}{{\rm d}}
\newcommand{\rR}{\mathbb R}
\newcommand{\code}[1]{\textbf{#1}}
\newcommand{\pkg}[1]{\textbf{#1}}
\let\proglang=\textit

\title{\pkg{simode}: \proglang{R} Package for statistical inference of ordinary differential equations using separable integral-matching}

\usepackage{authblk}

\author[1,2]{Rami Yaari \thanks{ramiyaari@gmail.com}}
\author[1]{Itai Dattner\thanks{idattner@stat.haifa.ac.il}}

\affil[1]{Department of Statistics, University of Haifa, Haifa, Israel}
\affil[2]{Bio-statistical and Bio-mathematical Unit, The Gertner Institute for Epidemiology and Health Policy Research, Chaim Sheba Medical Center, Tel Hashomer, Israel}

\begin{document}

\maketitle
\begin{abstract}
In this paper we describe \textbf{simode}: Separable Integral Matching for Ordinary Differential Equations. The statistical methodologies applied in the package focus on several minimization procedures of an integral-matching criterion function, taking advantage of the mathematical structure of the differential equations like separability of parameters from equations. Application of integral based methods to parameter estimation of ordinary differential equations was shown to yield more accurate and stable results comparing to derivative based ones. Linear features such as separability were shown to ease optimization and inference. We demonstrate the functionalities of the package using various systems of ordinary differential equations.
\end{abstract}





\section{Introduction}\label{sec:intro}
\subsection{Background}
This paper presents the \pkg{simode} \proglang{R} package \cite{R} aimed for conducting statistical inference on systems of ordinary differential equations (ODEs). Systems of ODEs are commonly used for the mathematical modeling of the rate of change of dynamic processes such as in mathematical biology \cite{edelstein2005mathematical}, biochemistry \cite{voit2000computational} and compartmental models in epidemiology \cite{anderson1992infectious}, to mention a few areas. Inference of ODEs involves the 'standard' statistical problems such as studying the identifiability of a model, estimating model parameters, predicting future states of the system, testing hypotheses, and choosing the 'best' model. However, dynamic systems are typically very complex: nonlinear, high dimensional and only partly measured. Moreover, data may be sparse and noisy. Thus, statistical learning (inference, prediction) of dynamical systems is not a trivial task in practice. In particular, numerical application of standard estimators, like the maximum likelihood or the least squares, may be difficult or computationally costly. Therefore, special computational platforms that allow for performing statistical inference for ODEs were recently developed. We first briefly mention some relevant packages we are aware of and then point out the main focus of \pkg{simode}.

Existing software implementations that are most relevant to this work are the following. \pkg{CollocInfer} \proglang{R} package of \cite{hooker2015collocinfer} implements the profiling methodology of \cite{ramsay2007parameter} and some extensions (there exist also a \proglang{Matlab} version). In the area of systems biology, \cite{raue2015data2dynamics} present \pkg{Data2Dynamics}, a modeling environment for \proglang{Matlab} that can be used for constructing dynamical models of biochemical reaction networks for large datasets and complex experimental conditions, and to perform efficient and reliable parameter estimation for model fitting. \cite{mikkelsen2017learning} developed the \pkg{episode} \proglang{R} package that implements adaptive integral-matching (AIM) algorithm for learning polynomial or rational ODEs with a sparse network structure. Other software libraries not directly related to our methodological framework are described or used in \cite{wood2010statistical}, \cite{wilkinson2011package}, \cite{JSSv069i12}, and \cite{wu2008dediscover} which focus on stochastic modeling or on more specific domains. \cite{voit2000computational} uses \proglang{PLAS} (Power Law Analysis and Simulation; Copyright 1996--2012 by Ant\'{o}nio Ferreira) a software suitable to analyze power-law differential equations. Also developed by Ant\'{o}nio Ferreira is \pkg{S-timator}, a \proglang{Python} library dedicated for analyzing ODE-based models.

The \proglang{R} package \pkg{simode} is substantially different from all the above tools in a sense that will be now explained and made clear.
\subsection[The focus of simode]{The focus of \pkg{simode}}
The statistical methodologies applied in the package are based on recent publications that study theoretical and applied aspects of smoothing methods in the context of ordinary differential equations (\cite{dattner2015}, \cite{dattner2015model}, \cite{dattner2017modelling}, \cite{yaarietal18}, \cite{dattnergugushvili18}). In that sense \pkg{simode} is closer in spirit to \pkg{CollocInfer} R package of \cite{hooker2015collocinfer}, and \pkg{episode} R package of \cite{mikkelsen2017learning}. Unlike \pkg{CollocInfer} we do not consider penalized estimation which balances between data and model. Further, we focus on integral-matching criterion functions which were shown to be more robust than gradient based ones \cite{dattner2015}. Using integral-matching criteria takes us closer in spirit to the \pkg{episode} R package of \cite{mikkelsen2017learning}. However, we focus on several minimization procedures of an integral-matching criterion function, taking advantage of the mathematical structure of the ODEs like separability of parameters from equations. Linear features such as separability were shown to ease optimization and inference (\cite{dattner2015}, \cite{dattner2015model}, \cite{vujavcic2015time}, \cite{dattner2017modelling}, \cite{yaarietal18}, \cite{vujavcic2018consistency}).

We demonstrate various functionalities of the package using different systems of ODEs. To be more specific, we demonstrate the ability of the package to implement a full estimation pipeline from point estimates to generating confidence intervals (using an example of S-system); deal with partially observed systems (using SIR example); user defined likelihood functions (using FitzHugh-Nagumo model); and the use of any function of 't' introduced into the ODE model. In addition, the package can implement Monte Carlo experiments sequentially or in parallel (demonstrated using the Lotka Volterra model).

The idea of separability of parameters and equations is now explained. Consider the following simple biochemical system taken from Chapter 2, Page 54 of \cite{voit2000computational},

\begin{equation}\label{eq:biosimple}
\begin{array}{l}
x_1^{\prime}(t)=2x_2(t)-1.2x_1(t)^{0.5}x_3(t)^{-1},
\\
x_2^{\prime}(t)=2x_1(t)^{0.1}x_3(t)^{-1}x_4(t)^{0.5}-2x_2(t),
\\
x_3=0.5,
\\
x_4=1.
\end{array}
\end{equation}

\begin{figure}[t!]
\centering
\includegraphics{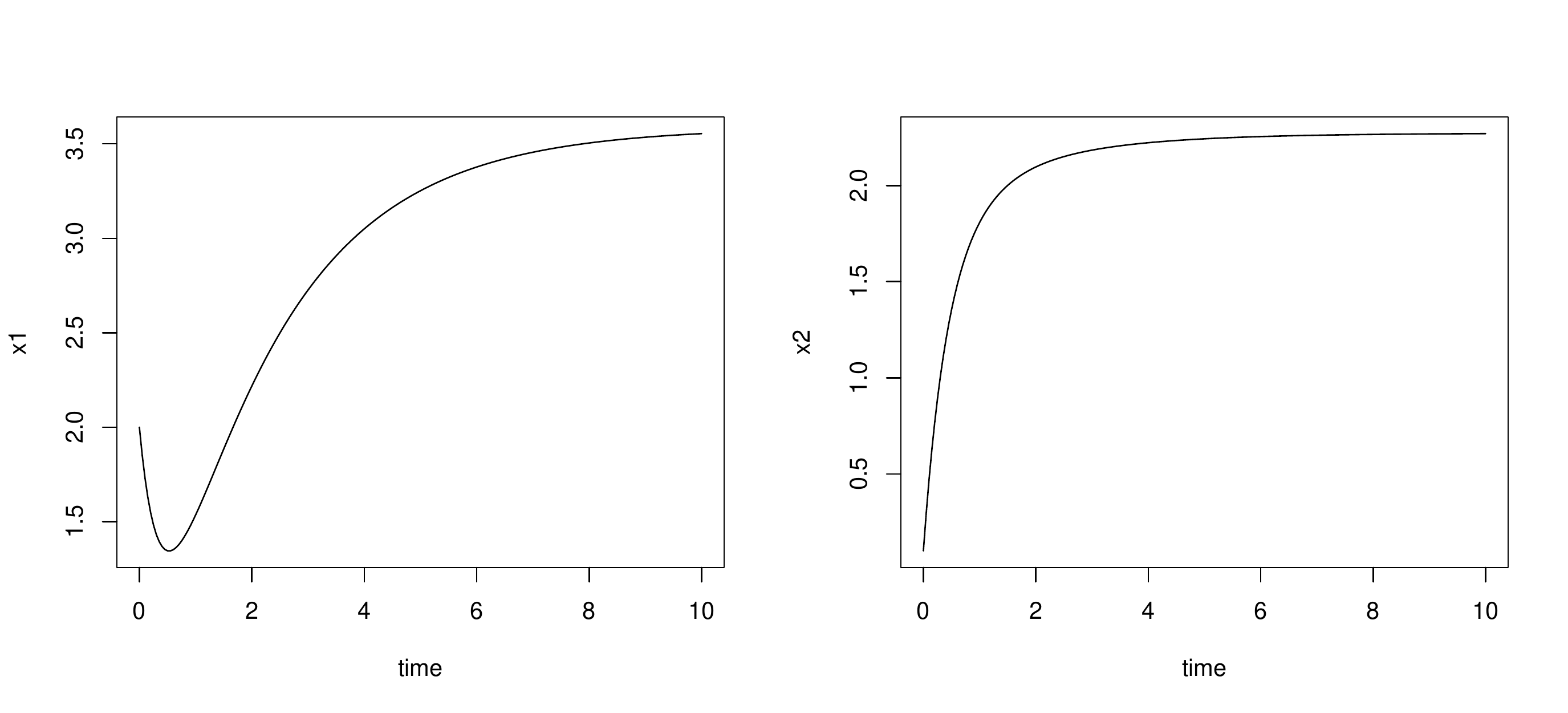}
\caption{\label{fig:simplebio} Solutions $x_1$ and $x_2$ of the biochemical system of equation (\ref{eq:biosimple}).}
\end{figure}

In this system the production of $x_2$ depends on $x_1,x_3$, and $x_4$ which enter with different kinetic orders (power). Specifically, $x_3$ has a negative power which indicates an inhibiting effect since an increase in $x_3$ leads to reduced production of $x_2$. The dynamics of the system for $x_1(0)=2$ and $x_2(0)=0.1$ is shown in Figure \ref{fig:simplebio}.

This system is a special case of an S-system (\cite{voit2000computational}) defined as
\begin{equation}\label{eq:s-system}
x^\prime_j(t)=\alpha_j\Pi_{k=1}^dx_k^{g_{jk}}(t)-\beta_j\Pi_{k=1}^dx_k^{h_{jk}}(t), \quad j=1\dots,d.
\end{equation}
Here, $\alpha_j,\beta_j$ are rate constants; $g_{jk},h_{jk}$ are kinetic orders that reflect the strength and directionality of the effect a variable has on a given influx or efflux. The above system is linear in  $\alpha_j,\beta_j$ but nonlinear in $g_{jk},h_{jk}$. In fact, one can view this system as a regression where the 'covariates' variables are $x_j(t)$, the solutions of the ODEs on the right hand side of the equations, while the 'response' variables are the derivatives $x_j^{\prime}(t)$ on the left hand side. Further, we can say that the system is linear in its rate constants but nonlinear in the kinetic orders (the powers).

More generally, consider a system of ordinary differential equations given by
\begin{equation}\label{eq:ode_model}
\bigg\{
\begin{array}{l}
 x^{\prime}(t)= F( x(t);\theta),\ t\in[0,T],
\\
 x(0)=\xi,
\end{array}
\end{equation}
where $ x(t)$ takes values in $\rR^d,\, \xi$ in $\Xi\subset \rR^d,$
and $\theta$ in $\Theta\subset\rR^p.$ The \pkg{simode} \proglang{R} package is especially useful for handling ODE systems for which
\begin{equation}\label{eq:sep_ode}
 F( x(t);\theta)= g(x(t);\theta_{NL})\theta_L,
\end{equation}
where $\theta=(\theta_{NL}^\top,\theta_{L}^\top)^\top$, $\top$ stands for the matrix transpose. Here $\theta_{NL}$, a vector of size $p_{NL}$, stands for the 'nonlinear' parameters that can not be separated from the state variables $x$, while $\theta_{L}$, a vector of size $p_L$, are the 'linear' parameters; note that $p=p_L+p_{NL}$. Setting \[\theta_{NL}=(g_{11},\dots,g_{1d},\dots,g_{d1},\dots,g_{dd},h_{11},\dots,h_{1d},\dots,h_{d1},\dots,h_{dd})^\top,\]
and $\theta_L=(\alpha_1,\beta_1,\dots,\alpha_d,\beta_d)^\top,$ one can easily see that (\ref{eq:s-system}) is a special case of (\ref{eq:ode_model}) with a vector field $F$ as given in (\ref{eq:sep_ode}). In the special case of the biochemical example (\ref{eq:biosimple}), fixing $x_3, x_4$, we have $d=2$, so that the matrix $g(x(t);\theta_{NL})$ is given by
\begin{eqnarray*}
\left(
\begin{array}{cccc}
x_1^{g_{11}}(t)x_2^{g_{12}}(t),-x_1^{h_{11}}(t)x_2^{h_{12}}(t),0,0
\\
0,0,x_1^{g_{21}}(t)x_2^{g_{22}}(t),-x_1^{h_{21}}(t)x_2^{h_{22}}(t)
\end{array}
\right).
\end{eqnarray*}
Thus, the system can be written in the form
\begin{eqnarray}\label{eq:biosimple_g}
\left(
\begin{array}{c}
x^\prime_1(t)
\\
x^\prime_2(t)
\end{array}
\right)
&=&
\left(
\begin{array}{cccc}
x_1^{g_{11}}(t)x_2^{g_{12}}(t),-x_1^{h_{11}}(t)x_2^{h_{12}}(t),0,0
\\
0,0,x_1^{g_{21}}(t)x_2^{g_{22}}(t),-x_1^{h_{21}}(t)x_2^{h_{22}}(t)
\end{array}
\right)\theta_L,
\end{eqnarray}
where $\theta_L=(\alpha_1,\beta_1,\alpha_2,\beta_2)^\top=(2,2.4,4,2)^\top$, \newline and $\theta_{NL}=(g_{11},g_{12},h_{11},h_{12},g_{21},g_{22},h_{21},h_{22})^\top=(0,1,0.5,0,0.1,0,0,1)^\top$. The vector field in the formulation above is separable in the linear parameter vector $\theta_L$, and therefore we refer to such systems as ODEs {\it linear in the parameter}  $\theta_L$ (as in a linear regression model). This linear property of the system turns out to be very useful for data fitting purposes where parameter estimation is required.

We emphasize that in our view \pkg{simode} should NOT be considered as a competitor for the other tools and packages mentioned above but instead, a complementary tool. Indeed, real problems arising in the area of dynamic systems are complex and typically there is no one method that can handle all type of problems uniformly better than other methods. For instance, one may consider generating initial parameter estimates using \pkg{simode} and then derive final estimates using generalized profiling implemented in \pkg{CollocInfer} that enables additional flexibility in assuming that the ODEs description of the dynamics are only approximately correct.

The paper is organized as follows. In the next section we briefly present the statistical methodology implemented in \pkg{simode}. In Section~\ref{sec:application} we describe in detail the use of \pkg{simode} for parameter estimation of ODEs. Section \ref{sec:part} deals with partial observed systems, while additional functionalities are demonstrated in Section \ref{sec:add}. The last section includes a summary and some future directions.

\section{Statistical methodology}\label{sec:stat_method}
Let $x(t; \theta,\xi), t \in [0, T ]$ be the solution of the initial value problem (\ref{eq:ode_model}) given values of $\xi$ and $\theta$. We assume measurements of $x$ are collected at discrete time points
\begin{equation}
\label{eq:obs}
Y_{j}(t_i)=x_j(t_i; \theta,\xi)+\epsilon_{ij}, \quad i=1,\ldots,n,j=1,\ldots,d,
\end{equation}
where the random variables $\epsilon_{ij}$ are independent
measurement errors (not necessarily Gaussian) with zero mean and finite variance. Consider the nonlinear least squares estimator of $\theta$ and $\xi$ defined as a minimiser with respect to $\eta$ and $\zeta$ of the least squares criterion function
\begin{eqnarray}\label{eq:nls}
\sum_{j=1}^d\sum_{i=1}^n (Y_{j}(t_i)-x_j(t_i; \eta,\zeta))^2.
\end{eqnarray}
Typically, the ODEs (\ref{eq:ode_model}) are nonlinear in $x$, and no analytic solution of the system exists, therefore numerical integration techniques are required in the estimation process. In fact, in the least squares criterion above the exact solution $x$ will be approximated by $\tilde{x}$, a numerical solution (e.g., using Runge-Kutta) of the ODEs equation (\ref{eq:ode_model}) for a given parameter and initial conditions. Thus, estimation methods such as nonlinear least squares or maximum likelihood require solving the system numerically for large set of potential parameters values, and then choosing an optimal parameter using some nonlinear optimisation technique. However, the combination of sparse and noisy data, nonlinear optimisation, and the need for numerical integration makes the parameter estimation a complex task (even for systems of low dimensions, e.g., \cite{voit2004decoupling}), and in many instances requires heavy computation. Therefore we adopt a 'smooth and match' approach for parameter estimation which includes two steps ({\it i}):  bypassing numerical integration by using nonparametric smoothing of the data, and ({\it ii}): estimating the parameters by fitting the ODEs model to the estimated functions (\cite{Bellman197126}, \cite{varah1982spline}, \cite{brunel2008parameter}, \cite{liang2008parameter}, \cite{fang2011two}, \cite{gugushvili2012sqrt}, \cite{dattner2015}, \cite{dattner2015model}, \cite{vujavcic2015time}, \cite{dattner2017modelling}, \cite{dattnergugushvili18}; see also Chapter 8 of \cite{ramsay2017dynamic}). Then the resulting estimates are used as initial guess for optimization of the nonlinear least squares criterion function (\ref{eq:nls}). In particular, in the first estimation stage which includes the two steps mentioned above, we consider integral-matching which is now described.
\subsection{Integral matching}
By integration, equation (\ref{eq:ode_model}) yields the system of integral equations
\begin{equation}\nonumber
x(t)=\xi + \int_0^t F( x(s);\theta)\, \rd s\,\ t\in[0,T].
\end{equation}
Here $x(t)=x(t;\theta,\xi)$ is the true solution of the ODE. Let $\hat{x}(t)$  stand for a nonparametric estimator (e.g., smoothing the data using splines or local polynomials) of the solution $x$ of the ODEs equation (\ref{eq:ode_model}) given observations (\ref{eq:obs}). The criterion function of an integral-matching approach for a fully observed systems of ODEs takes the form
\begin{equation}\label{eq:im}
\int_0^T\parallel\hat{x}(t)-\zeta - \int_0^t F( \hat x(s);\eta)\, \rd s\parallel^2 \rd t,
\end{equation}
where $\parallel \cdot \parallel$ denotes the standard Euclidean norm. The estimator of the parameter will be the minimiser of the criterion function (\ref{eq:im}), with respect to $\zeta$ and $\eta$. As its name suggests, integral-matching avoids the estimation of derivatives of the solution $x$ as done in other smooth and match applications and hence is more stable (\cite{dattner2015}). While applying integral-matching leads to stable estimators, minimizing the criterion (\ref{eq:im}) is a complicated task in practice and good initial guess of parameter values is required for optimization. Therefore, the \proglang{R} package \pkg{simode} is designed to take advantage of separability of the ODE system, as demonstrated above in equation (\ref{eq:biosimple_g}). This separability issue is now further explained.
\subsection{Exploiting linear features of the ODEs}
The \pkg{simode} package implements three separability scenarios corresponding to the following cases of equation (\ref{eq:ode_model}):
\begin{enumerate}
\item[(a)] ODEs linear in the parameters where $F( x(t);\theta)= g(x(t))\theta$;
\item[(b)] ODEs semi-linear in the parameters where $F( x(t);\theta)= g(x(t);\theta_{NL})\theta_L$;
\item[(c)] ODEs nonlinear in the parameters where $F( x(t);\theta)$ has no separable form that can be exploited.
\end{enumerate}

On top of the estimation stability the integral-matching criteria ensures, cases (a)-(b) above enable better optimization. Indeed, the above cases describe mathematical characteristics (separability) of the ODEs, and in what follows we use a smoothed version of separable nonlinear least squares (\cite{golub2003separable}) as well as 'classical' nonlinear least squares. These two optimization methods can be applied in both cases (a) and (b). However, case (c) characterizes a model for which the only optimization method applicable is a nonlinear one since there is no separability of parameters.

Consider case (a) of ODEs linear in the parameters where $F( x(t);\theta)= g(x(t))\theta$. Denote
\begin{eqnarray*}\label{eq:Ghat}
\hat{G}(t) &=& \int_0^t g(\hat{x}(s))\,\rd s\,,\quad t \in
[0,T],\nonumber \\
\hat{A} &=& \int_0^T \hat{G}(t)\,\rd t, \\
\hat{B} &=& \int_0^T \hat{G}^\top(t) \hat{G}(t)\,\rd t. \nonumber
\end{eqnarray*}

Minimizing the integral criterion function (\ref{eq:im}) with respect
to $\zeta$ and $\eta$ results in the direct estimators
\begin{eqnarray}
\hat{\xi} &=& \left(TI_d - \hat{A} \hat{B}^{-1}
\hat{A}^\top\right)^{-1} \int_0^T \left(I_d - \hat{A}
\hat{B}^{-1}
\hat{G}^\top(t)\right) \hat{x}(t)\,\rd t, \label{eq:xihat} \\
\hat{\theta}&=& \hat{B}^{-1} \int_0^T \hat{G}^\top(t) \left(
\hat{x}(t) -\hat{\xi} \right) \rd t, \label{eq:thetahat}
\end{eqnarray}
where $I_d$ denotes the $d \times d$ identity matrix. Note that
these estimators are well defined only if the inverse matrices in
(\ref{eq:xihat}) and (\ref{eq:thetahat}) exist. Necessary and sufficient conditions for $\sqrt{n}$-consistency of the 'direct integral estimators' (\ref{eq:xihat}) and (\ref{eq:thetahat}) are provided in \cite{dattner2015}. Furthermore, the extensive simulation study in the aforementioned paper has demonstrated that using integrals as above instead of derivatives yields more accurate estimates. Indeed, it is well known (see e.g.,  \cite{voit2000computational} and \cite{chou2009recent}) that estimating derivatives from noisy and sparse data may be rather inaccurate. Additional  application of the direct integral method to a variety of synthetic and real data was shown to yield accurate and stable results in \cite{vujavcic2015time} and \cite{dattnergugushvili18}. Clearly, in this special case of ODEs linear in the parameter $\theta$ the complex task of nonlinear optimization reduces to the least squares solutions (\ref{eq:xihat}) and (\ref{eq:thetahat}) which are easy to obtain and therefore a substantial computational improvement in optimization performance is achieved.

Now consider case (b) above of ODEs semi-linear in the parameters for which in equation (\ref{eq:ode_model}) the model can be written in the form $F( x(t);\theta)= g(x(t);\theta_{NL})\theta_L$. Then, for a given $\theta_{NL}$, minimizing the integral criterion function (\ref{eq:im}) yields least squares solutions similar to (\ref{eq:xihat})-(\ref{eq:thetahat}) which we denote by $\hat{\xi}(\theta_{NL})$ and $\hat{\theta}_L(\theta_{NL})$ with the notation emphasizing the dependence of the linear solutions on the nonlinear parameters. Plugging back $\hat{\xi}(\theta_{NL})$ and $\hat{\theta}_L(\theta_{NL})$ into the integral criterion function (\ref{eq:im}) results with
\begin{equation}\label{eq:sepnls}
M(\theta_{NL}):=\int_0^T\left|\left|
\hat{x}(t)-\hat{\xi}(\theta_{NL})-\hat{G}(t;\theta_{NL})\hat{\theta}_L(\theta_{NL})
\right|\right|^2\rd t,
\end{equation}
where we have defined $\hat{G}(t;\theta_{NL})= \int_0^t g(\hat{x}(s);\theta_{NL})\,\rd s\,,\quad t \in
[0,T]\nonumber$. Once $M(\theta_{NL})$ is minimized and a solution $\hat\theta_{NL}$ is obtained, estimators for $\xi$ and $\theta$ follow immediately and are given by $\hat{\xi}(\hat\theta_{NL})$ and ${\hat\theta}_L(\hat\theta_{NL})$, respectively. This optimization procedure was considered in \cite{dattner2017modelling} and is a form of separable nonlinear least squares (\cite{golub2003separable}). Note that the we apply nonlinear optimization only for estimating the nonlinear parameters $\theta_{NL}$, and hence the dimension of the optimization problem has been substantially reduced.

Finally, case (c) above requires nonlinear optimization for estimating $\xi$ and $\theta$ and the dimension of the optimization problem can not be reduced.
\section[Parameter estimation of ODEs using simode]{Parameter estimation of ODEs using \pkg{simode}}\label{sec:application}
In this section we demonstrate the main functionality of \pkg{simode} package (version 1.0.0), for parameter estimation of ODEs, via exploring the cases
\begin{enumerate}
\item[(a)] ODEs linear in the parameters where $F( x(t);\theta)= g(x(t))\theta$;
\item[(b)] ODEs semi-linear in the parameters where $F( x(t);\theta)= g(x(t);\theta_{NL})\theta_L$;
\item[(c)] ODEs nonlinear in the parameters where $F( x(t);\theta)$ has no separable form that can be exploited.
\end{enumerate}
We continue with the biochemical example described in equation (\ref{eq:biosimple}). Consider equation (\ref{eq:biosimple_g}) where we fix $x_3=0.5, x_4=1$. Further, assume that the zero kinetic parameters are known , so that the system is given by

\begin{equation}\label{eq:biosimple2}
\begin{array}{l}
x_1^{\prime}(t)=2x_2(t)-2.4x_1(t)^{0.5},
\\
x_2^{\prime}(t)=4x_1(t)^{0.1}-2x_2(t).
\end{array}
\end{equation}

\begin{figure}[t!]
\centering
\includegraphics{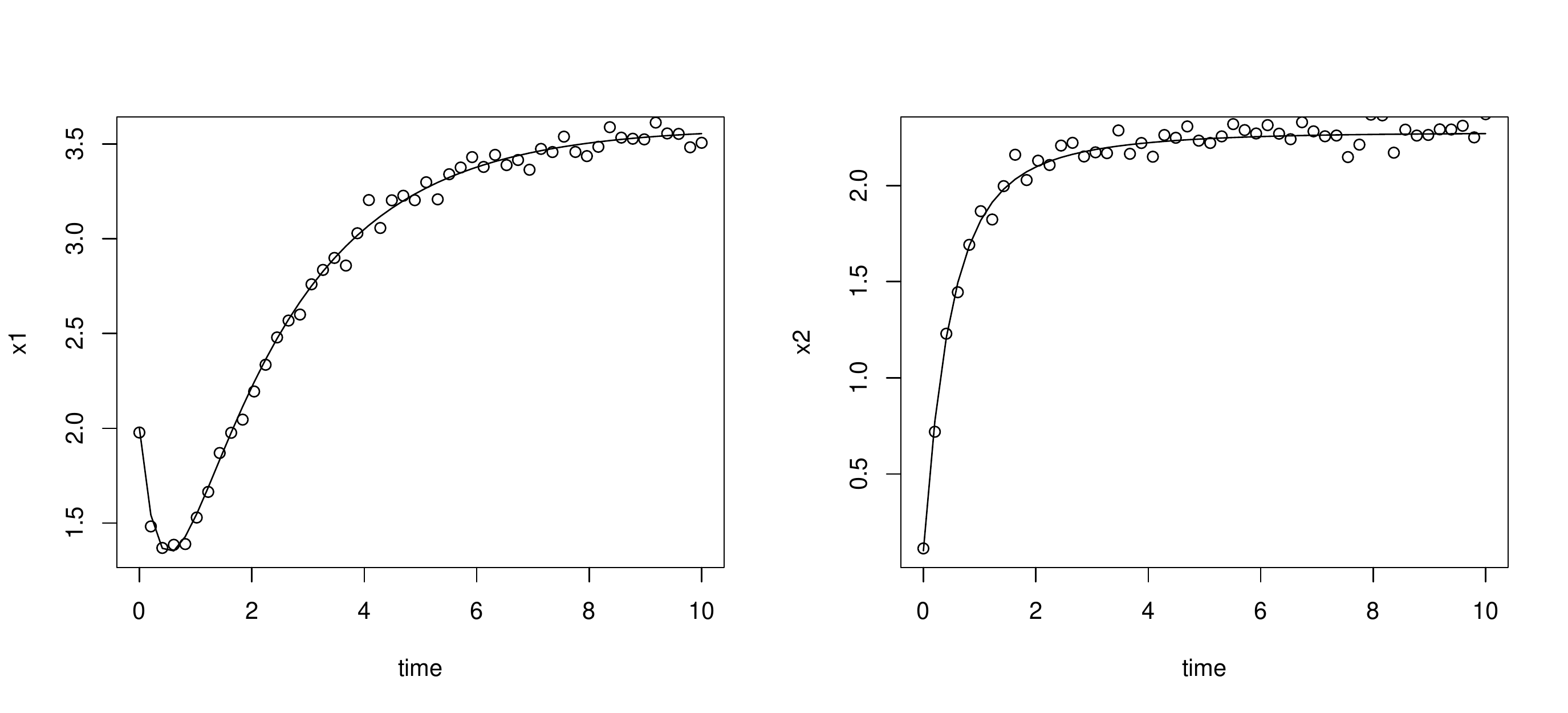}
\caption{\label{fig:simplebio2} Solutions $x_1$ and $x_2$ of the biochemical system of equation (\ref{eq:biosimple}) in solid lines. Stochastic observations generated from the statistical model (\ref{eq:obs}) in circles.}
\end{figure}

Thus, the linear parameters of the system are $\theta_L=(\alpha_1,\beta_1,\alpha_2,\beta_2)^\top=(2,2.4,4,2)^\top$, and the nonlinear parameters are $\theta_{NL}=(g_{12},h_{11},g_{21},h_{22})^\top=(1,0.5,0.1,1)^\top$. Here is the system given in equation (\ref{eq:biosimple2}) as it should be written to be used later by the package:

\begin{Schunk}
\begin{Sinput}
R> pars <- c('alpha1','g12','beta1','h11', 'alpha2','g21','beta2','h22')
R> vars <- paste0('x', 1:2)
R> eq1 <- 'alpha1*(x2^g12)-beta1*(x1^h11)'
R> eq2 <- 'alpha2*(x1^g21)-beta2*(x2^h22)'
R> equations <- c(eq1,eq2)
R> names(equations) <- vars
R> theta <- c(2,1,2.4,0.5,4,0.1,2,1)
R> names(theta) <- pars
R> x0 <- c(2,0.1)
R> names(x0) <- vars
\end{Sinput}
\end{Schunk}
The code above is the symbolic setup of the ODE system. The resulting objects of variables, parameters and initial conditions are
\begin{Schunk}
\begin{Sinput}
R> equations
\end{Sinput}
\begin{Soutput}
                              x1                               x2 
"alpha1*(x2^g12)-beta1*(x1^h11)" "alpha2*(x1^g21)-beta2*(x2^h22)" 
\end{Soutput}
\begin{Sinput}
R> theta
\end{Sinput}
\begin{Soutput}
alpha1    g12  beta1    h11 alpha2    g21  beta2    h22 
   2.0    1.0    2.4    0.5    4.0    0.1    2.0    1.0 
\end{Soutput}
\begin{Sinput}
R> x0
\end{Sinput}
\begin{Soutput}
 x1  x2 
2.0 0.1 
\end{Soutput}
\end{Schunk}
Since we are working with symbolic objects we have created a function \code{solve\_ode} that uses the \code{ode} function of \pkg{deSolve} package (\cite{deSolve}). The following code generates observations according to the statistical model defined in equation (\ref{eq:obs}) where the distribution of the measurement error is Gaussian with standard deviation of $0.05$. The resulting 'true' ODE solutions and the stochastic observations are presented in Figure (\ref{fig:simplebio2}).

\begin{Schunk}
\begin{Sinput}
R> library("simode")
R> set.seed(1000)
R> n <- 50
R> time <- seq(0,10,length.out=n)
R> model_out <- solve_ode(equations,theta,x0,time)
R> x_det <- model_out[,vars]
R> sigma <- 0.05
R> obs <- list()
R> for(i in 1:length(vars)) {
+    obs[[i]] <- x_det[,i] + rnorm(n,0,sigma)
+  }
R> names(obs) <- vars
\end{Sinput}
\end{Schunk}

Now that we have setup the system of ODEs in a symbolic form and generated observations from the statistical model we can explore cases (a)-(b)-(c) described above: we estimate model parameters, plot model fits, and provide profile-likelihood confidence intervals.

The package uses integral-matching as a first stage and then (by default) executes nonlinear least squares  optimization, namely minimizing equation (\ref{eq:nls}), starting from the integral-matching estimates. The first estimation stage, i.e., the integral-matching, is based on smoothing the observations. We use by default the \code{smooth.spline} method of the \pkg{stats} package, with generalized cross validation (we also support performing integral-matching without smoothing the observations). 
Our implementation of the estimators (\ref{eq:xihat})-(\ref{eq:thetahat}) uses \code{lsqlincon} function of the \pkg{pracma} package, which also allows us to introduce constraints on the parameters.

\subsection{Case (a): ODEs linear in the parameters}
For simplicity we begin with assuming that the initial conditions $x_1(0), x_2(0)$ are known to the user. Here we also assume that the kinetic parameters are all known, so that our goal is to estimate the vector $\theta_L=(\alpha_1,\beta_1,\alpha_2,\beta_2)^\top$. The code for doing so is:
\begin{Schunk}
\begin{Sinput}
R> lin_pars <- c('alpha1','beta1','alpha2','beta2')
R> nlin_pars <- setdiff(pars,lin_pars)
R> est_lin <- simode(equations=equations, pars=lin_pars, 
+    fixed=c(x0,theta[nlin_pars]), time=time, obs=obs)
R> summary(est_lin)
\end{Sinput}
\begin{Soutput}
call:
simode(equations = equations, pars = lin_pars, time = time, 
    obs = obs, fixed = c(x0, theta[nlin_pars]))

equations:
                            x1                             x2 
"alpha1*(x2^1)-beta1*(x1^0.5)" "alpha2*(x1^0.1)-beta2*(x2^1)" 

initial conditions:
 x1  x2 
2.0 0.1 

parameter estimates:
     par   type im_est nls_est
1 alpha1 linear  1.932   2.013
2  beta1 linear  2.324   2.432
3 alpha2 linear  3.868   3.943
4  beta2 linear  1.923   1.959

im-method:  separable 

im-loss:  0.1492 

nls-loss:  0.2398 
\end{Soutput}
\end{Schunk}
The call to \code{simode} returns a \code{simode} object containing the parameters estimates obtained using integral-matching as well as those obtained using nonlinear least squares optimization starting from the integral-matching estimates.

\begin{figure}[t!]
\centering
\begin{Schunk}
\begin{Sinput}
R> plot(est_lin, type='fit', pars_true=theta[lin_pars],
+       mfrow=c(1,2),legend=T)
\end{Sinput}
\end{Schunk}
\includegraphics{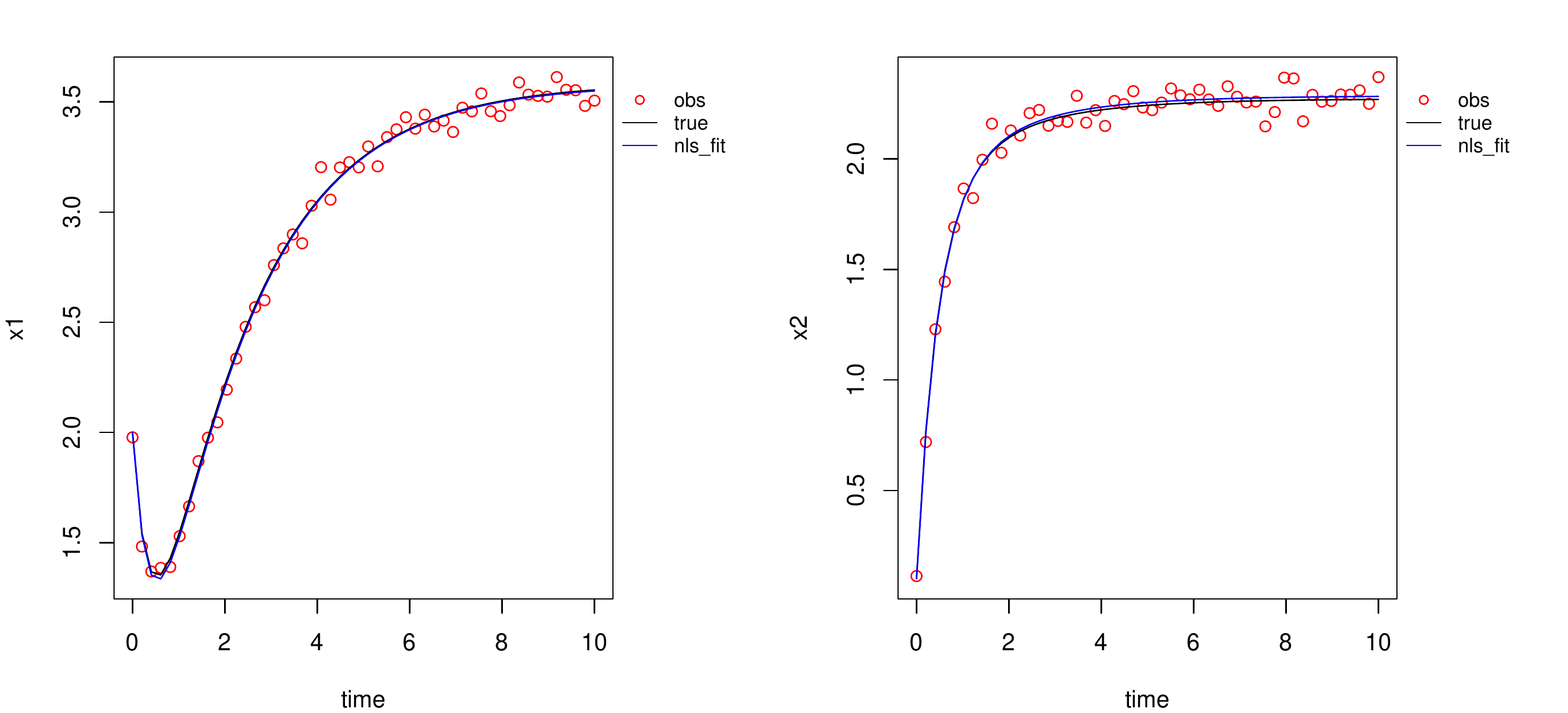}
\caption{\label{fig:simplebio3} Case (a): 'True' and estimated solutions $x_1$ and $x_2$ of the biochemical system of equation (\ref{eq:biosimple}).}
\end{figure}

An implementation of the generic plot function for \code{simode} objects can be used to plot the fit obtained using these estimates, either the fit after
integral-matching and nonlinear least squares optimization (the default), just the integral-matching based fit, or both; see Figure \ref{fig:simplebio3}. In this case, we can also plot the fit against the true curves since we know the true values of the parameters that were used to generate the observations. The same plot function can also be used to show the estimates obtained as demonstrated in the sequel.

In the code above we have defined which parameters are linear by \code{lin\_pars} and which are not by \code{nlin\_pars}. Then we fixed the set of parameters and initial conditions that we do not want to estimate. However, note that it is not mandatory for the user to know which parameters are linear and which are not. For instance, here is the result of running the estimation without this knowledge:
\begin{Schunk}
\begin{Sinput}
R> est_par <- simode(
+    equations=equations,time=time,pars=pars,fixed=x0,obs=obs)
\end{Sinput}
\begin{Soutput}
Problem in eq.1 [x1] - parameter [g12] should be set as non-linear
Problem in eq.1 [x1] - parameter [h11] should be set as non-linear
Problem in eq.2 [x2] - parameter [g21] should be set as non-linear
Problem in eq.2 [x2] - parameter [h22] should be set as non-linear
\end{Soutput}
\end{Schunk}
As can be seen, the \code{simode} function generates error messages that point out exactly the nonlinear parameters. This is due to the fact that the \code{simode} function considers ODEs linear in the parameters as its default, namely \code{im\_method = "separable"}, unless \code{im\_method = "non-separable"} is defined. This added \code{simode} feature makes it very useful for handling ODEs with linear features in case the mathematical knowledge for characterizing them is lacking.

\begin{figure}[t!]
\centering
\begin{Schunk}
\begin{Sinput}
R> plot(profile_lin, mfrow=c(2,2))
\end{Sinput}
\end{Schunk}
\includegraphics{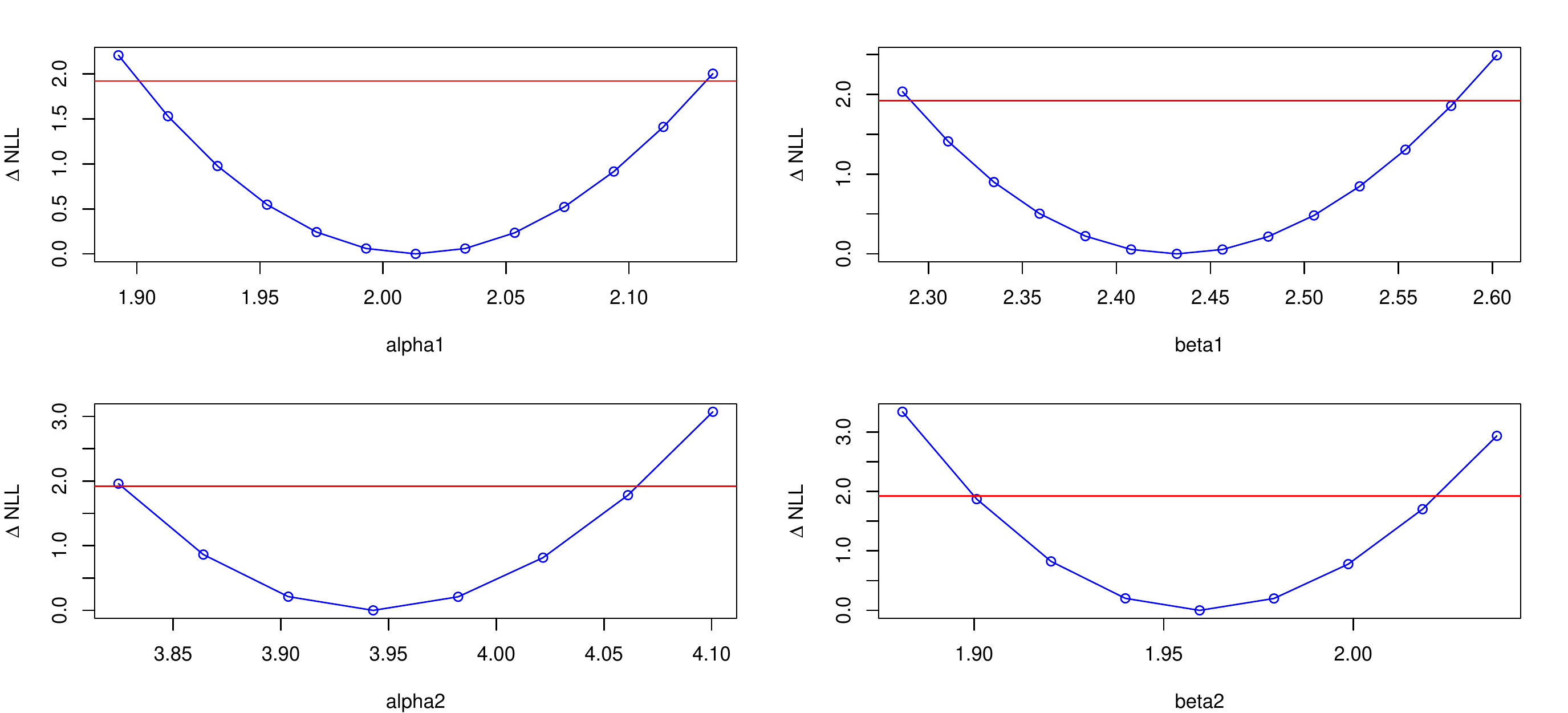}
\caption{\label{fig:profile_lin} Profile likelihood confidence intervals for the linear parameters of the system (\ref{eq:biosimple}).}
\end{figure}

Now we generate and plot confidence intervals for the parameters using profile likelihood, see Figure \ref{fig:profile_lin}. In case nonlinear optimization for the point estimates was used, then the profiling is done using a Gaussian based likelihood with fixed sigma which we estimate in the background.

\begin{Schunk}
\begin{Sinput}
R> step_size <- 0.01*est_lin$nls_pars_est
R> profile_lin <- profile(est_lin,step_size=step_size)
R> confint(profile_lin,level=0.95)
\end{Sinput}
\begin{Soutput}
call:
confint.profile.simode(object = profile_lin, level = 0.95)
level:
0.95 
intervals:
     par  nls_est    lower    upper
1 alpha1 2.013303 1.901046 2.130440
2  beta1 2.432117 2.290981 2.581607
3 alpha2 3.942877 3.825985 4.065744
4  beta2 1.959493 1.899323 2.021247
\end{Soutput}
\end{Schunk}
\subsection{Case (b): ODEs semi-linear in the parameters}
Now consider ODEs semi-linear in the parameters where separable nonlinear least squares might be used as in equation (\ref{eq:sepnls}). Thus, the nonlinear parameters $\theta_{NL}=(g_{12},h_{11},g_{21},h_{22})^\top=(1,0.5,0.1,1)^\top$ are not assumed to be known and their estimation is needed. Estimating nonlinear parameters requires nonlinear optimization. The function \code{simode} uses the \code{optim} function, thus we need to provide initial guess for optimization. In our example, the true parameter values are known, so we generate random initial guess in the vicinity of the true nonlinear parameters. The code and estimation results are given below.

\begin{figure}[t!]
\centering
\begin{Schunk}
\begin{Sinput}
R> plot(est_semilin, type='est', show='both', pars_true=theta, legend=T)
\end{Sinput}
\end{Schunk}
\includegraphics{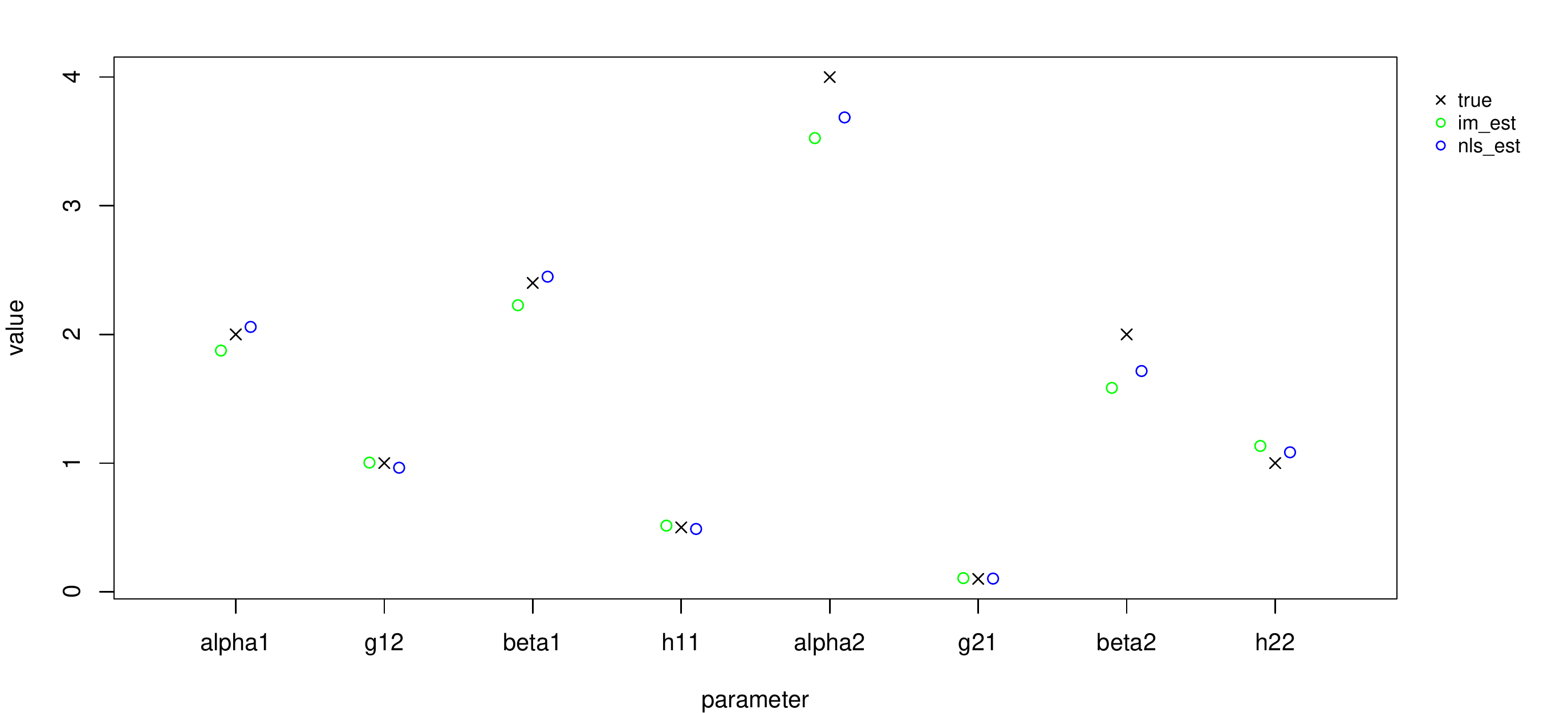}
\caption{\label{fig:caseb_pars} Case (b) separable: 'True' and estimated parameters of the biochemical system of equation (\ref{eq:biosimple}).}
\end{figure}

\begin{Schunk}
\begin{Sinput}
R> nlin_init <- rnorm(length(theta[nlin_pars]),theta[nlin_pars],
+                      0.1*theta[nlin_pars])
R> names(nlin_init) <- nlin_pars
R> est_semilin <- simode(
+    equations=equations, pars=pars, fixed=x0, time=time, obs=obs,
+    nlin_pars=nlin_pars, start=nlin_init)
R> summary(est_semilin)
\end{Sinput}
\begin{Soutput}
call:
simode(equations = equations, pars = pars, time = time, obs = obs, 
    nlin_pars = nlin_pars, fixed = x0, start = nlin_init)

equations:
                              x1                               x2 
"alpha1*(x2^g12)-beta1*(x1^h11)" "alpha2*(x1^g21)-beta2*(x2^h22)" 

initial conditions:
 x1  x2 
2.0 0.1 

parameter estimates:
     par       type      start im_est nls_est
1 alpha1     linear         NA 1.8740  2.0580
2    g12 non-linear 0.86305878 1.0040  0.9637
3  beta1     linear         NA 2.2270  2.4490
4    h11 non-linear 0.50815084 0.5136  0.4877
5 alpha2     linear         NA 3.5260  3.6870
6    g21 non-linear 0.09886774 0.1057  0.1019
7  beta2     linear         NA 1.5840  1.7160
8    h22 non-linear 1.08597553 1.1330  1.0840

im-method:  separable 

im-loss:  0.1142 

nls-loss:  0.239 
\end{Soutput}
\end{Schunk}
We can plot the resulting integral-maching and nonlinear least squares parameter estimates and compare them visualy to their true values, see Figure \ref{fig:caseb_pars}.

So far we have used the default optimization of \code{simode} function which executes separable least squares. However, we could ignore the fact that the ODE we consider has linear features in its parameters. In that case we execute classical nonlinear optimization of the integral-matching criterion function for all the parameters, $\theta_L$, and $\theta_{NL}$.

To run \code{simode} in non-separable mode set the argument \code{im\_method} accordingly. Initial guesses for the nonlinear parameters are still obligatory. Setting initial guesses for the linear parameters is optional in this case. However, unless specifically entered, initial guesses for the optimization used in order to estimate the linear parameters appearing in the integral-matching criterion function are calculated directly using the separability of the model.

\begin{Schunk}
\begin{Sinput}
R> est_semilin_nosep <- simode(
+    equations=equations, pars=pars, fixed=x0,
+    time=time, obs=obs, nlin_pars=nlin_pars, start=nlin_init,
+    im_method = "non-separable")
R> summary(est_semilin_nosep)
\end{Sinput}
\begin{Soutput}
call:
simode(equations = equations, pars = pars, time = time, obs = obs, 
    nlin_pars = nlin_pars, fixed = x0, start = nlin_init, 
    im_method = "non-separable")

equations:
                              x1                               x2 
"alpha1*(x2^g12)-beta1*(x1^h11)" "alpha2*(x1^g21)-beta2*(x2^h22)" 

initial conditions:
 x1  x2 
2.0 0.1 

parameter estimates:
     par       type      start  im_est nls_est
1 alpha1     linear         NA 1.76300  2.0690
2    g12 non-linear 0.86305878 1.05000  0.9597
3  beta1     linear         NA 2.12500  2.4580
4    h11 non-linear 0.50815084 0.53230  0.4862
5 alpha2     linear         NA 3.63800  3.6930
6    g21 non-linear 0.09886774 0.09619  0.1013
7  beta2     linear         NA 1.70000  1.7230
8    h22 non-linear 1.08597553 1.07100  1.0800

im-method:  non-separable 

im-loss:  0.1146 

nls-loss:  0.2389 
\end{Soutput}
\end{Schunk}

\begin{figure}[t!]
\centering
\begin{Schunk}
\begin{Sinput}
R> plot(est_semilin_nosep, type='fit', pars_true=theta, 
		mfrow=c(1,2), legend=T)
\end{Sinput}
\end{Schunk}
\includegraphics{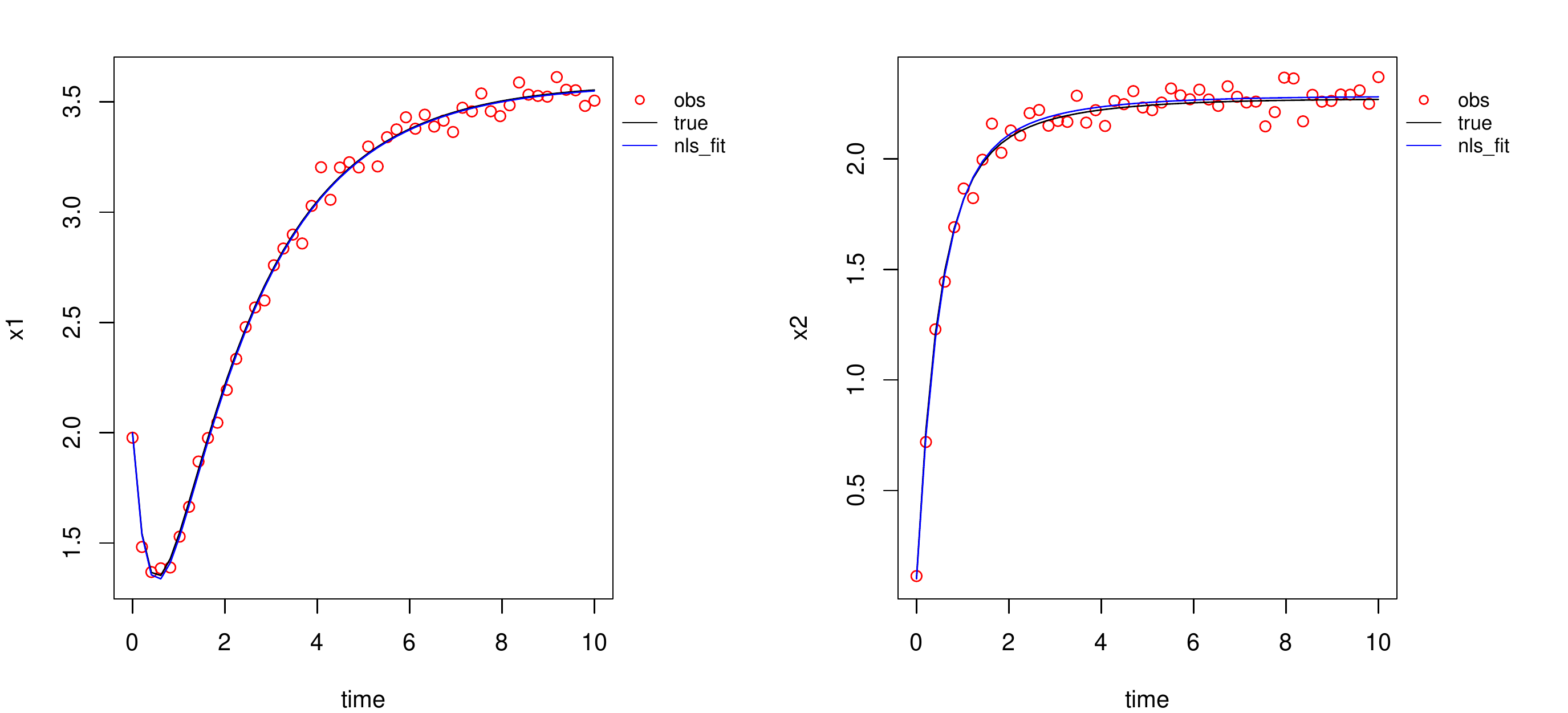}
\caption{\label{fig:casebnosep} Case (b) non-separable: 'True' and estimated solutions $x_1$ and $x_2$ of the biochemical system of equation (\ref{eq:biosimple}).}
\end{figure}

\subsection{Case (c): ODEs nonlinear in the parameters}
There are cases where the system of ODEs is not separable, meaning that there are only nonlinear parameters. For instance, consider our toy example where the linear parameters, namely the coefficients are known. Still we may want to use integral-matching since this way we bypass the need for solving numerically the ODEs, at least in the first stage of optimization. Doing so leads to minimization of (\ref{eq:im}) as it is. The resulting code is:

\begin{Schunk}
\begin{Sinput}
R> est_nosep <- simode(
+    equations=equations, pars=pars, nlin_pars=pars, 
+    start=nlin_init, fixed=c(theta[lin_pars],x0),
+    im_method = 'non-separable', time=time, obs=obs)
R> summary(est_nosep)
\end{Sinput}
\begin{Soutput}
call:
simode(equations = equations, pars = pars, time = time, 
    obs = obs, nlin_pars = pars, fixed = c(theta[lin_pars], x0), 
    start = nlin_init, im_method = "non-separable")

equations:
                       x1                        x2 
"2*(x2^g12)-2.4*(x1^h11)"   "4*(x1^g21)-2*(x2^h22)" 

initial conditions:
 x1  x2 
2.0 0.1 

parameter estimates:
  par       type      start im_est nls_est
1 g12 non-linear 0.86305878 0.8303 0.97600
2 h11 non-linear 0.50815084 0.3975 0.48830
3 g21 non-linear 0.09886774 0.4040 0.08095
4 h22 non-linear 1.08597553 1.4490 0.96560

im-method:  non-separable 

im-loss:  38.3 

nls-loss:  0.2401 
\end{Soutput}
\end{Schunk}

\subsection{Initial conditions $x(0)$ are unknown}\label{sub:initunknown}
In the examples above, for simplicity of presentation, we considered the initial conditions to be known. Here is an example of estimating the initial conditions using the separability property of the ODEs as in case (b) (separable) above. We add the names of the unknown $x_0$ variables to the list of parameters to estimate.

\begin{Schunk}
\begin{Sinput}
R> est_all <- simode(
+    equations=equations, pars=c(pars,names(x0)), time=time, 
+    obs=obs, nlin_pars=nlin_pars, start=nlin_init)
R> summary(est_all)
\end{Sinput}
\begin{Soutput}
call:
simode(equations = equations, pars = c(pars, names(x0)), time = time, 
    obs = obs, nlin_pars = nlin_pars, start = nlin_init)

equations:
                              x1                               x2 
"alpha1*(x2^g12)-beta1*(x1^h11)" "alpha2*(x1^g21)-beta2*(x2^h22)" 

initial conditions:
x1 x2 
NA NA 

parameter estimates:
      par       type      start im_est nls_est
1  alpha1     linear         NA 1.4350  1.7800
2     g12 non-linear 0.86305878 1.1610  1.0480
3   beta1     linear         NA 1.7340  2.1450
4     h11 non-linear 0.50815084 0.6020  0.5327
5  alpha2     linear         NA 3.2440  3.6790
6     g21 non-linear 0.09886774 0.1204  0.1023
7   beta2     linear         NA 1.3320  1.7090
8     h22 non-linear 1.08597553 1.2650  1.0870
9      x1     linear         NA 1.9190  1.9580
10     x2     linear         NA 0.1282  0.1037

im-method:  separable 

im-loss:  0.106 

nls-loss:  0.2374 
\end{Soutput}
\end{Schunk}

Note that unless otherwise defined, the \code{simode} method implements the estimator for initial conditions defined in (\ref{eq:xihat}). Alternatively, one could estimate the initial conditions using nonlinear optimization by adding the initial conditions to the \code{nlin\_pars} argument in the call to \code{simode}. However, in that case an intial guess $x_0$ for optimization is required.
\section{Partially observed systems of ODEs}\label{sec:part}

In general, inference using integral-matching requires a fully observed system. However, in some cases, integral-matching can be applied to a partially observed system. For example, if it is possible to reconstruct the unobserved variables using estimates of the system parameters. We demonstrate this using an example of an ODE system describing the spread of seasonal influenza in multiple age-groups across mutiple seasons. A discrete-time version of the model and a two-stage estimation procedure similar to the one used
in \pkg{simode}, was described in details in \cite{yaarietal18}. Here we present the model and show how to employ the \pkg{simode} package in order to estimate its parameters.

The model is an SIR-type (Susceptible-Infected-Recovered) model. The epidemic
in each age-group $1\leq a \leq M$ and each season $1\leq y \leq L$ can be described using two equations for the proportion of susceptible ($S$) and infected ($I$) in the population (the proportion of recovered is given by $1-S-I$):

\begin{equation}\label{eq:sir}
\begin{array}{l}
S_{a,y}^{\prime}(t)=-S_{a,y}(t)\kappa_y\sum_{j=1}^{M}\beta_{a,j}I_{j,y}(t),
\\
I_{a,y}^{\prime}(t)=S_{a,y}(t)\kappa_y\sum_{j=1}^{M}(\beta_{a,j}I_{j,y}(t))-\gamma I_{a,y}(t).
\end{array}
\end{equation}

The parameters of the model include the $M\times M$ transmission matrix $\beta$, the recovery rate $\gamma$ and $\kappa_{2,...,L}$ which signify the relative infectivity of the influenza virus strains circulating in seasons $2,...,L$ compared to season $1$ ($\kappa_1$ is fixed as $1$). As shown in \cite{yaarietal18}, taking into account separability characteristics of this model is advantageous.

The \pkg{simode} package includes an example dataset called \code{sir\_example}, containing pre-made structures for testing this example. In this dataset there are two age-groups and five influenza seasons, so in total there are $10$ equations for $S$ and $10$ equations for $I$.

\begin{Schunk}
\begin{Sinput}
R> data(sir_example)
R> summary(sir_example)
\end{Sinput}
\begin{Soutput}
          Length Class  Mode     
equations 20     -none- character
beta       4     -none- numeric  
gamma      1     -none- numeric  
kappa      4     -none- numeric  
S0        10     -none- numeric  
I0        10     -none- numeric  
time      18     -none- numeric  
obs       10     -none- list     
\end{Soutput}
\begin{Sinput}
R> sir_example$beta
\end{Sinput}
\begin{Soutput}
beta1_1 beta2_1 beta1_2 beta2_2 
      6       2       1       3 
\end{Soutput}
\begin{Sinput}
R> sir_example$gamma
\end{Sinput}
\begin{Soutput}
   gamma 
2.333333 
\end{Soutput}
\begin{Sinput}
R> sir_example$kappa
\end{Sinput}
\begin{Soutput}
kappa2 kappa3 kappa4 kappa5 
 0.988  1.182  1.037  1.052 
\end{Soutput}
\begin{Sinput}
R> sir_example$S0
\end{Sinput}
\begin{Soutput}
S1_1 S2_1 S1_2 S2_2 S1_3 S2_3 S1_4 S2_4 S1_5 S2_5 
0.56 0.57 0.49 0.45 0.56 0.32 0.56 0.47 0.47 0.41 
\end{Soutput}
\end{Schunk}

The dataset contains noisy observations for the $I$ variables created using the parameter and initial condition values given in the example according to model (\ref{eq:sir}), where the measurement errors have a Gaussian distribution with $\sigma=0.001$. There are no observations of the $S$ variables in the example as the proportion of susceptible in the population is typically unknown. Time is given in weeks and include $18$ weeks of observations. The values of the parameters $\beta$ and $\gamma$ are also given assuming a time unit of weeks.

\subsection{Case (a): SIR linear in the parameters}

We begin exploring this example in the simplest case in which we assume all
the parameter values besides the matrix $\beta$, as well as all the initial conditions are known. However,  the integral-matching method requires a fully observed system while in this case there are no observations for the $S$ variables. Nevertheless, given the observations of the $I$ variables, and given values for the parameter $\gamma$ and the initial conditions $S_{a,y}(0)$, $I_{a,y}(0)$, we can generate observations for the $S$ variables using the formula:

$$S_{a,y}(t)=S_{a,y}(0)+I_{a,y}(0)-I_{a,y}(t)-\gamma \int_{u=1}^{t} I_{a,y}(u)du$$

\begin{figure}[t!]
\centering
\begin{Schunk}
\begin{Sinput}
R> plot(est_sir_lin, type='fit', which=names(I0),
+       time=seq(1,time[length(time)],by=0.1),
+       pars_true=beta, mfrow=c(5,2))
\end{Sinput}
\end{Schunk}
\includegraphics{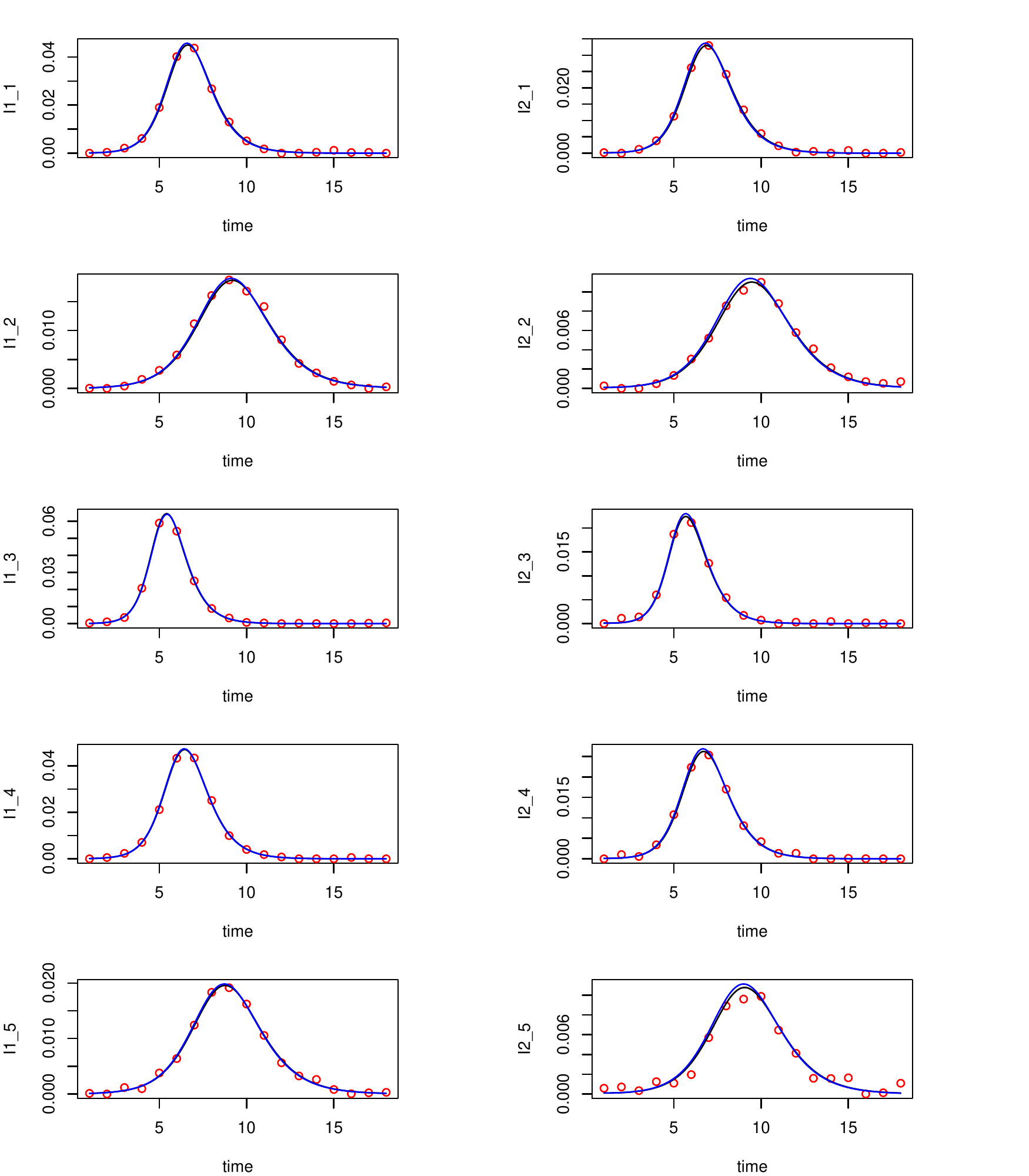}
\caption{\label{fig:sir_a1} SIR case (a) - fit to observations of $I$}
\end{figure}

\begin{figure}[t!]
\centering
\begin{Schunk}
\begin{Sinput}
R> plot(est_sir_lin, type='est', show='both',
+       pars_true=beta, legend=T)
\end{Sinput}
\end{Schunk}
\includegraphics{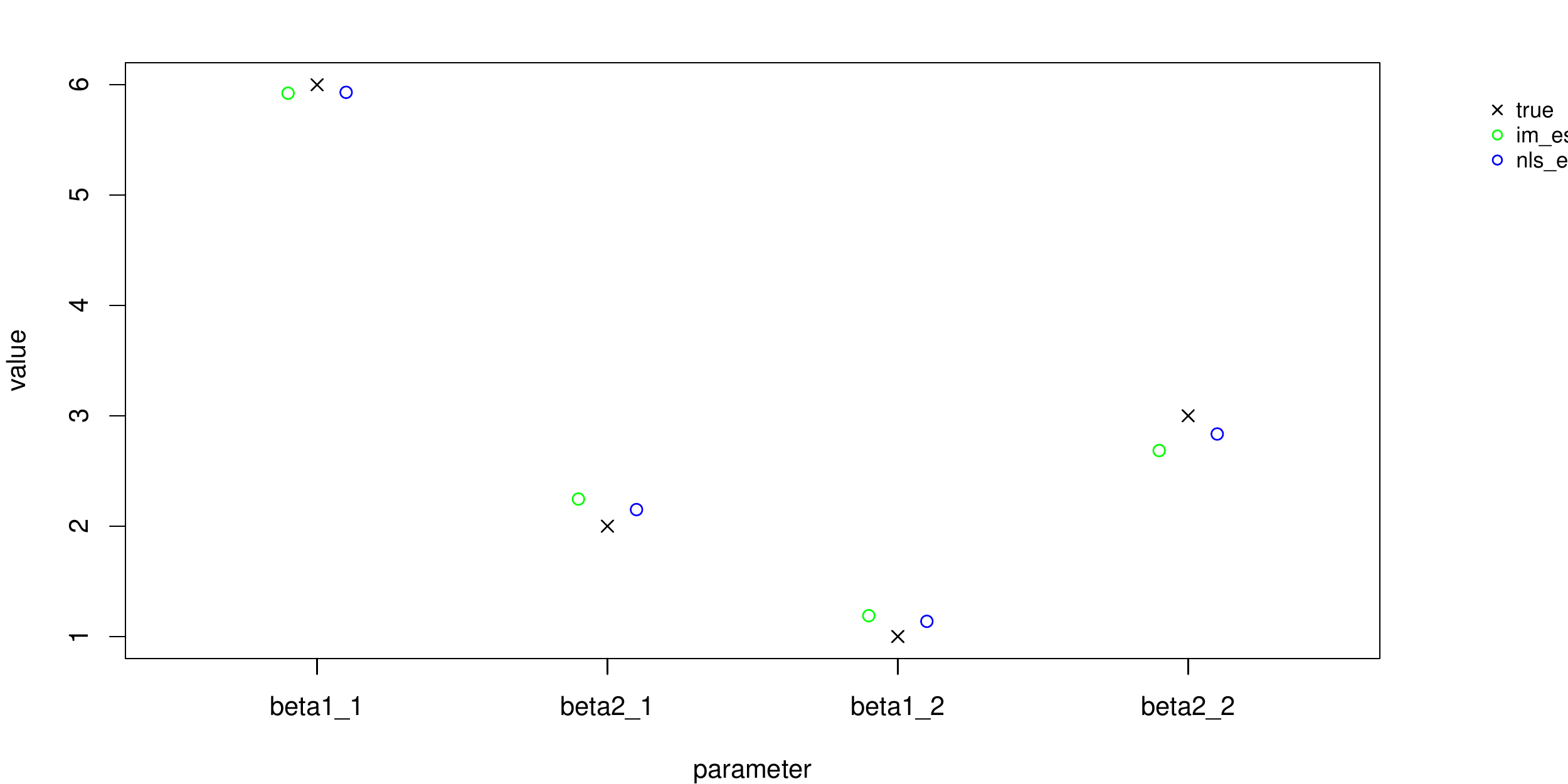}
\caption{\label{fig:sir_a2} SIR case (a) - estimates of $\beta$}
\end{figure}

In the code below, observations for the $S$ variables are generated using the above formula. We then fit the data assuming the parameters $\gamma$ and $\kappa$ and the initial conditions are known. 
Resulting data fits and parameter estimates are presented in Figures \ref{fig:sir_a1}-\ref{fig:sir_a2}.

\begin{Schunk}
\begin{Sinput}
R> equations <- sir_example$equations
R> S0 <- sir_example$S0
R> I0 <- sir_example$I0
R> beta <- sir_example$beta
R> gamma <- sir_example$gamma
R> kappa <- sir_example$kappa
R> time <- sir_example$time
R> I_obs <- sir_example$obs
R> S_obs <- lapply(1:length(S0),function(j)
+      S0[j] + I0[j] - I_obs[[j]]
+      - gamma*pracma::cumtrapz(time,I_obs[[j]]))
R> names(S_obs) <- names(S0)
R> obs <- c(S_obs,I_obs)
R> x0 <- c(S0,I0)
R> pars <- names(beta)
R> pars_min <- rep(0,length(pars))
R> names(pars_min) <- pars
R> est_sir_lin <- simode(
+    equations=equations, pars=pars, time=time, obs=obs,
+    fixed=c(gamma,kappa,x0), lower=pars_min)
R> summary(est_sir_lin)$est
\end{Sinput}
\begin{Soutput}
      par   type lower im_est nls_est
1 beta1_1 linear     0  5.924   5.932
2 beta2_1 linear     0  2.246   2.150
3 beta1_2 linear     0  1.188   1.137
4 beta2_2 linear     0  2.685   2.835
\end{Soutput}
\end{Schunk}

\subsection[Case (b1): SIR - semi-linear with unknown initial conditions - gamma and kappa are known]{Case (b1): SIR - semi-linear with unknown initial conditions - $\gamma$ and $\kappa$ are known}

Now let us assume that the initial conditions $S_0$ are not known. In this case, we cannot generate the missing observations for the susceptible variables ahead of time. However, if we run \code{simode} in order to estimate $\beta$ and $S_0$, then for a given set of values of $S_0$ within nonlinear optimization, we can generate the missing observations and estimate $\beta$ using integral-matching. To do this, we need to define a function that will generate the missing observations given the missing parameters, and pass this function in the call to \code{simode} using the \code{gen\_obs} argument. The user-defined \code{gen\_obs} function should accept as arguments the equations, parameter values, initial condition values, time points and observations. Additional arguments could be passed to \code{gen\_obs} by passing them in the call to \code{simode} using the ellipsis construct. The function should return a list with two elements: the element 'obs' which should contain the observations for all the equations in the ODE, and the element 'time' with the time points for the observations (since 'time' can include different time points for each equation). In this example, we define a function that receives additional parameters that include $\gamma$ and the names of the $S$ and $I$ variables. Here is the code for this case followed by estimation results.

\begin{Schunk}
\begin{Sinput}
R> gen_obs <- function(equations, pars, x0, time, obs,
+                      gamma, S_names, I_names, ...)
+  {
+    S0 <- x0[S_names]
+    I0 <- x0[I_names]
+    I_obs <- obs
+    S_obs <- lapply(1:length(S0),function(i)
+      S0[i]+I0[i]-I_obs[[i]]-gamma*pracma::cumtrapz(time,I_obs[[i]]))
+    names(S_obs) <- S_names
+    obs <- c(S_obs,I_obs)
+    return (list(obs=obs, time=time))
+  }
R> pars <- c(names(beta),names(S0))
R> pars_min <- rep(0,length(pars))
R> names(pars_min) <- pars
R> pars_max <- rep(1,length(S0))
R> names(pars_max) <- names(S0)
R> S0_init <- rep(0.5,length(S0))
R> names(S0_init) <- names(S0)
R> est_sir_semilin <- simode(
+    equations=equations, pars=pars, time=time, obs=I_obs,
+    fixed=c(I0,gamma,kappa), nlin=names(S0), start=S0_init,
+    lower=pars_min, upper=pars_max,  gen_obs=gen_obs,
+    gamma=gamma, S_names=names(S0), I_names=names(I0))
R> summary(est_sir_semilin)$est
\end{Sinput}
\begin{Soutput}
       par       type lower upper start  im_est nls_est
1  beta1_1     linear     0   Inf    NA 6.47300  6.3400
2  beta2_1     linear     0   Inf    NA 1.91200  1.9720
3  beta1_2     linear     0   Inf    NA 0.02678  0.0000
4  beta2_2     linear     0   Inf    NA 2.99100  2.9920
5     S1_1 non-linear     0     1   0.5 0.57560  0.5847
6     S2_1 non-linear     0     1   0.5 0.58360  0.5733
7     S1_2 non-linear     0     1   0.5 0.49930  0.5046
8     S2_2 non-linear     0     1   0.5 0.45660  0.4537
9     S1_3 non-linear     0     1   0.5 0.55370  0.5575
10    S2_3 non-linear     0     1   0.5 0.34100  0.3279
11    S1_4 non-linear     0     1   0.5 0.56410  0.5710
12    S2_4 non-linear     0     1   0.5 0.48700  0.4740
13    S1_5 non-linear     0     1   0.5 0.47330  0.4811
14    S2_5 non-linear     0     1   0.5 0.40880  0.3979
\end{Soutput}
\end{Schunk}

\subsection[Case (b2): SIR - semi-linear with unknown initial conditions - gamma and kappa are unknown]{Case (b2): SIR - semi-linear with unknown initial conditions - $\gamma$ and $\kappa$ are unknown}

Finally, we can try and estimate all the model parameters including the nonlinear parameters $\gamma$ and $\kappa_{2,...,5}$. We define the function
\code{gen\_obs2} in which the value of the parameter $\gamma$ is taken from the \code{pars} argument.

\begin{Schunk}
\begin{Sinput}
R> gen_obs2 <- function(equations, pars, x0, time, obs,
+                       S_names, I_names, ...)
+  {
+    gamma <- pars['gamma']
+    S0 <- x0[S_names]
+    I0 <- x0[I_names]
+    I_obs <- obs
+    S_obs <- lapply(1:length(S0),function(i)
+      S0[i]+I0[i]-I_obs[[i]]-gamma*pracma::cumtrapz(time,I_obs[[i]]))
+    names(S_obs) <- S_names
+    obs <- c(S_obs,I_obs)
+    return (list(obs=obs, time=time))
+  }
R> gamma_init <- 2
R> names(gamma_init) <- names(gamma)
R> kappa_init <- rep(1,length(kappa))
R> names(kappa_init) <- names(kappa)
R> pars <- names(c(beta,gamma,kappa,S0))
R> nlin_pars <- names(c(gamma,kappa,S0))
R> start <- c(gamma_init,kappa_init,S0_init)
R> names(start) <- nlin_pars
R> pars_min <- c(rep(0,length(beta)),1.4,rep(0.25,length(kappa)),
+                rep(0,length(S0)))
R> pars_max <- c(rep(Inf,length(beta)),3.5,rep(4,length(kappa)),
+                rep(1,length(S0)))
R> names(pars_min) <- pars
R> names(pars_max) <- pars
R> est_sir_all <- simode(
+    equations=equations, pars=pars, time=time, obs=I_obs,
+    nlin_pars=nlin_pars, start=start, fixed=I0,
+    lower=pars_min, upper=pars_max,
+    gen_obs=gen_obs2, S_names=names(S0), I_names=names(I0))
R> summary(est_sir_all)$est
\end{Sinput}
\begin{Soutput}
       par       type lower upper start  im_est nls_est
1  beta1_1     linear  0.00   Inf    NA 6.60300 6.54900
2  beta2_1     linear  0.00   Inf    NA 1.75200 1.78200
3  beta1_2     linear  0.00   Inf    NA 0.05009 0.04751
4  beta2_2     linear  0.00   Inf    NA 3.85800 3.85200
5    gamma non-linear  1.40   3.5   2.0 2.33100 2.25700
6   kappa2 non-linear  0.25   4.0   1.0 0.89450 0.97130
7   kappa3 non-linear  0.25   4.0   1.0 1.15500 1.16800
8   kappa4 non-linear  0.25   4.0   1.0 1.05600 1.03200
9   kappa5 non-linear  0.25   4.0   1.0 1.00100 1.05400
10    S1_1 non-linear  0.00   1.0   0.5 0.56690 0.55240
11    S2_1 non-linear  0.00   1.0   0.5 0.53850 0.51820
12    S1_2 non-linear  0.00   1.0   0.5 0.52900 0.48350
13    S2_2 non-linear  0.00   1.0   0.5 0.46070 0.41450
14    S1_3 non-linear  0.00   1.0   0.5 0.55370 0.53560
15    S2_3 non-linear  0.00   1.0   0.5 0.33330 0.31070
16    S1_4 non-linear  0.00   1.0   0.5 0.54960 0.54290
17    S2_4 non-linear  0.00   1.0   0.5 0.44910 0.43600
18    S1_5 non-linear  0.00   1.0   0.5 0.48280 0.45280
19    S2_5 non-linear  0.00   1.0   0.5 0.39780 0.36290
\end{Soutput}
\end{Schunk}
\section{Additional functionalities of the package}\label{sec:add}

In this section we demonstrate some additional functionalities of the package. We do that using other systems of ODEs in order to further explore the package usability.

\subsection{User defined likelihood function}

Consider the case where the user has her own likelihood function to be used in the second stage of optimization, meaning after the integral-matching stage. The default optimization of the package implements in the second stage the nonlinear least squares loss function (\ref{eq:nls}). This default estimation procedure suffices for the method to result in consistent estimators. However, one may prefer to implement a specific likelihood function, for example a Gaussian distribution with known or unknown variance. We demonstrate this option using as an example the FitzHugh-Nagumo spike potential equations where the ODE model is given by
\begin{equation}\label{eq:fn}
\begin{array}{l}
V^{\prime}(t)=c(V(t)-V(t)^{3}/3+R(t)),
\\
R^{\prime}(t)=-(V(t)-a+bR(t))/c.
\end{array}
\end{equation}
See \cite{hooker2015collocinfer} for further explanation of this model. The above system is linear in $a, b$ but nonlinear in $c$. The following code sets the equations, parameters and the 'true' values for initial conditions and parameters:
\begin{Schunk}
\begin{Sinput}
R> pars <- c('a','b','c')
R> vars <- c('V','R')
R> eq_V <- 'c*(V-V^3/3+R)'
R> eq_R <- '-(V-a+b*R)/c'
R> equations <- c(eq_V,eq_R)
R> names(equations) <- vars
R> x0 <- c(-1,1)
R> names(x0) <- vars
R> theta <- c(0.2,0.2,3)
R> names(theta) <- pars
\end{Sinput}
\end{Schunk}
The following code generates observations from a Gaussian measurement error model:
\begin{Schunk}
\begin{Sinput}
R> n <- 40
R> time <- seq(0,20,length.out=n)
R> model_out <- solve_ode(equations,theta,x0,time)
R> x_det <- model_out[,vars]
R> set.seed(1000)
R> sigma <- 0.05
R> obs <- list()
R> for(i in 1:length(vars)) {
+    obs[[i]] <- x_det[,i] + rnorm(n,0,sigma)
+  }
R> names(obs) <- vars
\end{Sinput}
\end{Schunk}

Here we implement a Gaussian distribution (negative-log) likelihood function,
which will be passed in the call to \code{simode} using the \code{calc\_nll} argument. The function should receive as arguments the parameter values, time points, observations and output of the solutions of the ODEs (calculated using the estimated parameter values in the current iteration of the optimization). Additional arguments can be passed to \code{calc\_nll} by passing them in the call to \code{simode} using the ellipsis construct (in this case the parameter \code{sigma} is passed as well). Note that the user-defined likelihood will also be used in the call to profile and the calculation of confidence intervals based on the likelihood profiles. Here is the user defined (negative-log) Gaussian likelihood function,

\begin{Schunk}
\begin{Sinput}
R> calc_nll <- function(pars, time, obs, model_out, sigma, ...) {
+    -sum(unlist(lapply(names(obs),function(var) {
+      dnorm(obs[[var]],mean=model_out[,var],sd=sigma,log=T)
+    })))
+  }
\end{Sinput}
\end{Schunk}

\begin{figure}[t!]
\centering
\begin{Schunk}
\begin{Sinput}
R> plot(est_fn, type='fit', pars_true=theta, mfrow=c(2,1), legend=T)
\end{Sinput}
\end{Schunk}
\includegraphics{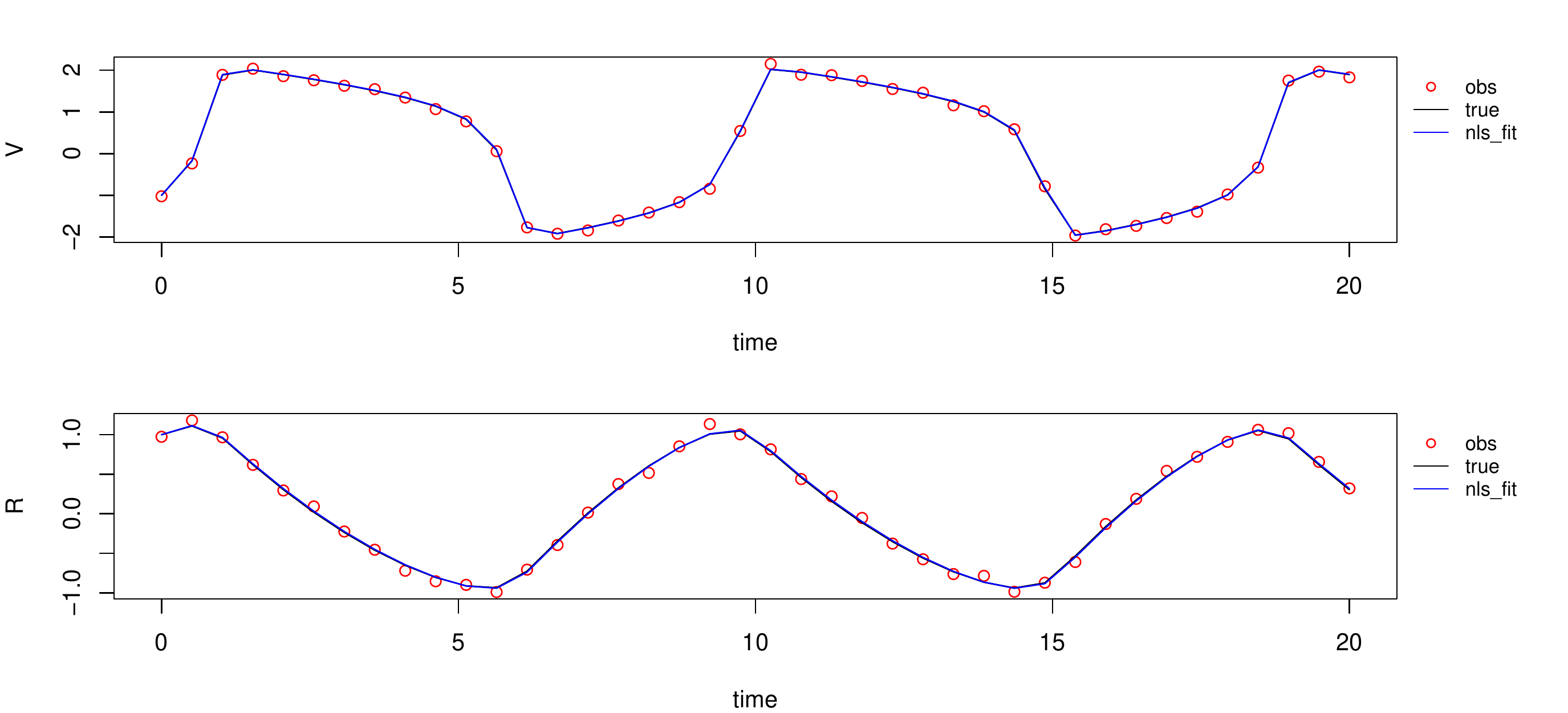}
\caption{\label{fig:fn} Solutions $V$ and $R$ of the FitzHugh-Nagum model (\ref{eq:fn}), noisy observations and model fits.}
\end{figure}

Now we demonstrate the usage of the likelihood function defined above, where the variance is assumed to be known and therefore is given as a fixed parameter. In this example the nonlinear parameters are assumed to be known. The resulting model fits are presented in Figure \ref{fig:fn}. The parameter estimates are

\begin{Schunk}
\begin{Sinput}
R> lin_pars <- c('a','b')
R> nlin_pars <- c('c')
R> init_vals <-  rnorm(length(theta[nlin_pars]),
+   		theta[nlin_pars],0.1*theta[nlin_pars])
R> names(init_vals) <- nlin_pars
R> est_fn <- simode(
+    equations=equations, pars=pars, time=time, obs=obs,
+    fixed=x0, nlin_pars=nlin_pars, start=init_vals,
+    calc_nll=calc_nll, sigma=sigma)
R> summary(est_fn)
\end{Sinput}
\begin{Soutput}
call:
simode(equations = equations, pars = pars, time = time, obs = obs, 
    nlin_pars = nlin_pars, fixed = x0, start = init_vals, 
    calc_nll = calc_nll, sigma = sigma)

equations:
              V               R 
"c*(V-V^3/3+R)"  "-(V-a+b*R)/c" 

initial conditions:
 V  R 
-1  1 

parameter estimates:
  par       type    start im_est nls_est
1   a     linear       NA 0.1724  0.2040
2   b     linear       NA 0.2365  0.1767
3   c non-linear 3.350783 3.3050  3.0020

im-method:  separable 

im-loss:  6.815 

nls-loss:  -130.9 
\end{Soutput}
\end{Schunk}

Now the above example is explored but with unknown variance which requires estimation as well. The user likelihood function has to be slightly modified as follows.

\begin{Schunk}
\begin{Sinput}
R> calc_nll_sig <- function(pars, time, obs, model_out, ...) {
+    sigma <- pars['sigma']
+    -sum(unlist(lapply(names(obs),function(var) {
+      dnorm(obs[[var]],mean=model_out[,var],sd=sigma,log=T)
+    })))
+  }
R> names(sigma) <- 'sigma'
R> lik_pars <- names(sigma)
R> pars_fn_sig <- c(pars,lik_pars)
R> init_vals[names(sigma)] <- 0.3
R> lower <- NULL
R> lower[names(sigma)] <- 0
R> est_fn_sig <- simode(
+    equations=equations, pars=pars_fn_sig, time=time, obs=obs,
+    fixed=x0, nlin_pars=nlin_pars, likelihood_pars=lik_pars,
+    start=init_vals, lower=lower, calc_nll=calc_nll_sig)
R> summary(est_fn_sig)$est
\end{Sinput}
\begin{Soutput}
    par       type lower    start im_est nls_est
1     a     linear  -Inf       NA 0.1724 0.20400
2     b     linear  -Inf       NA 0.2365 0.17670
3     c non-linear  -Inf 3.350783 3.3050 3.00200
4 sigma likelihood     0 0.300000     NA 0.04693
\end{Soutput}
\end{Schunk}

\subsection{System Decoupling}
\cite{voit2004decoupling} demonsrated that system decoupling combined with data smoothing may lead to better reconstruction of the underlying dynamic system, and to better estimation of parameters. We implemented this functionality within the \code{simode} package. We use the ODE system of the previous example (\ref{eq:fn}) for demonstrating decoupling functionality. The only difference is that we now set \code{decouple\_equations=T} in the \code{simode} function. The resulting integral-matching estimated values are stored in the returned \code{simode} object (as usual). If there is a parameter shared across system equations, the \code{simode} function will use the mean value (across system equations) of these parameter estimates (parameter 'c' in this example). The parameter estimates for each equation before averaging are stored in the \code{simode} object in a matrix called \code{im\_pars\_est\_mat}. In this matrix, parameters that are not part of a given equation are set with NA values.

\begin{Schunk}
\begin{Sinput}
R> init_vals <- init_vals[nlin_pars]
R> est_fn_d <- simode(
+    equations=equations, pars=pars, time=time, obs=obs,
+    fixed=x0, nlin_pars=nlin_pars, start=init_vals,
+    calc_nll=calc_nll, sigma=sigma, decouple_equations=T)
R> est_fn_d$im_pars_est_mat
\end{Sinput}
\begin{Soutput}
          a         b        c
V        NA        NA 1.686992
R 0.1945346 0.1707976 2.910466
\end{Soutput}
\begin{Sinput}
R> summary(est_fn_d)$est
\end{Sinput}
\begin{Soutput}
  par       type    start im_est nls_est
1   a     linear       NA 0.1945  0.2040
2   b     linear       NA 0.1708  0.1767
3   c non-linear 3.350783 2.2990  3.0020
\end{Soutput}
\end{Schunk}
\subsection{Monte Carlo simulations}

Conducting Monte Carlo experiments for ODEs can be an intensive computational task. The \code{simode} function can be given multiple sets of observations
of a system from Monte Carlo simulations and fit each of these sets separately, returning a list of \code{simode} objects with the parameter estimates obtained from each fit. We demonstrate this using as an example the predator-prey Lotka-Volterra model (\cite{edelstein2005mathematical}) given by the equations ($X$=prey, $Y$=predator):

\begin{equation}\label{eq:lv}
\begin{array}{l}
X^{\prime}(t)=\alpha X(t)-\beta X(t) Y(t),
\\
Y^{\prime}(t)=\delta X(t)Y(t)-\gamma Y(t).
\end{array}
\end{equation}

The Lotka-Volterra system is linear in all of its parameters. The following code sets the equations, parameters and the 'true' values for intial conditions and parameters:
\begin{Schunk}
\begin{Sinput}
R> pars <- c('alpha','beta','gamma','delta')
R> vars <- c('X','Y')
R> eq_X <- 'alpha*X-beta*X*Y'
R> eq_Y <- 'delta*X*Y-gamma*Y'
R> equations <- c(eq_X,eq_Y)
R> names(equations) <- vars
R> x0 <- c(0.9,0.9)
R> names(x0) <- vars
R> theta <- c(2/3,4/3,1,1)
R> names(theta) <- pars
\end{Sinput}
\end{Schunk}
Next, we generate ten Monte Carlo sets of observations from a Gaussian measurement error model.

\begin{Schunk}
\begin{Sinput}
R> n <- 100
R> time <- seq(0,25,length.out=n)
R> model_out <- solve_ode(equations,theta,x0,time)
R> x_det <- model_out[,vars]
R> N <- 10
R> mc_obs <- list()
R> sigma <- 0.1
R> set.seed(1000)
R> for(j in 1:N) {
+    obs <- list()
+    for(i in 1:length(vars)) {
+      obs[[i]] <- rnorm(n,x_det[,i],sigma)
+    }
+    names(obs) <- vars
+    mc_obs[[j]] <- obs
+  }
\end{Sinput}
\end{Schunk}

To fit the ten sets of Monte Carlo simulations in one call to \code{simode},
simply set \code{obs} to the list with the Monte Carlo observations (\code{mc\_obs} in this case) and set the argument \code{obs\_sets} to the number of Monte Carlo sets. We set the control parameter \code{obs\_sets\_fit} in \code{simode.control} to 'separate' (the default) to indicate we want to fit each observation set separately. By default, the sets will be fitted sequentially. To fit them in parallel, set the control parameter \code{parallel} in \code{simode.control} to true:

\begin{Schunk}
\begin{Sinput}
R> lv_mc <- simode(
+    equations=equations, pars=c(pars,vars), time=time, obs=mc_obs,
+    obs_sets=N, simode_ctrl=simode.control(parallel=T))
R> summary(lv_mc,sum_mean_sd=T,pars_true=c(theta,x0),digits=2)$est
\end{Sinput}
\begin{Soutput}
    par true im_mean im_sd im_bias im_rmse nls_mean nls_sd nls_bias
1 alpha 0.67    0.65 0.039  -0.017   0.042     0.68 0.0230    0.013
2  beta 1.30    1.30 0.073  -0.033   0.081     1.40 0.0400    0.067
3 gamma 1.00    0.92 0.030  -0.080   0.081     0.98 0.0340   -0.020
4 delta 1.00    0.93 0.032  -0.070   0.080     0.98 0.0340   -0.020
5     X 0.90    0.84 0.047  -0.060   0.075     0.91 0.0290    0.010
6     Y 0.90    0.86 0.050  -0.040   0.061     0.90 0.0089    0.000
  nls_rmse
1   0.0270
2   0.0530
3   0.0380
4   0.0370
5   0.0310
6   0.0086
\end{Soutput}
\end{Schunk}

\begin{figure}[t!]
\centering
\begin{Schunk}
\begin{Sinput}
R> plot(lv_mc, type='est', show='both', plot_mean_sd=T,
+       pars_true=c(theta,x0), legend=T)
\end{Sinput}
\end{Schunk}
\includegraphics{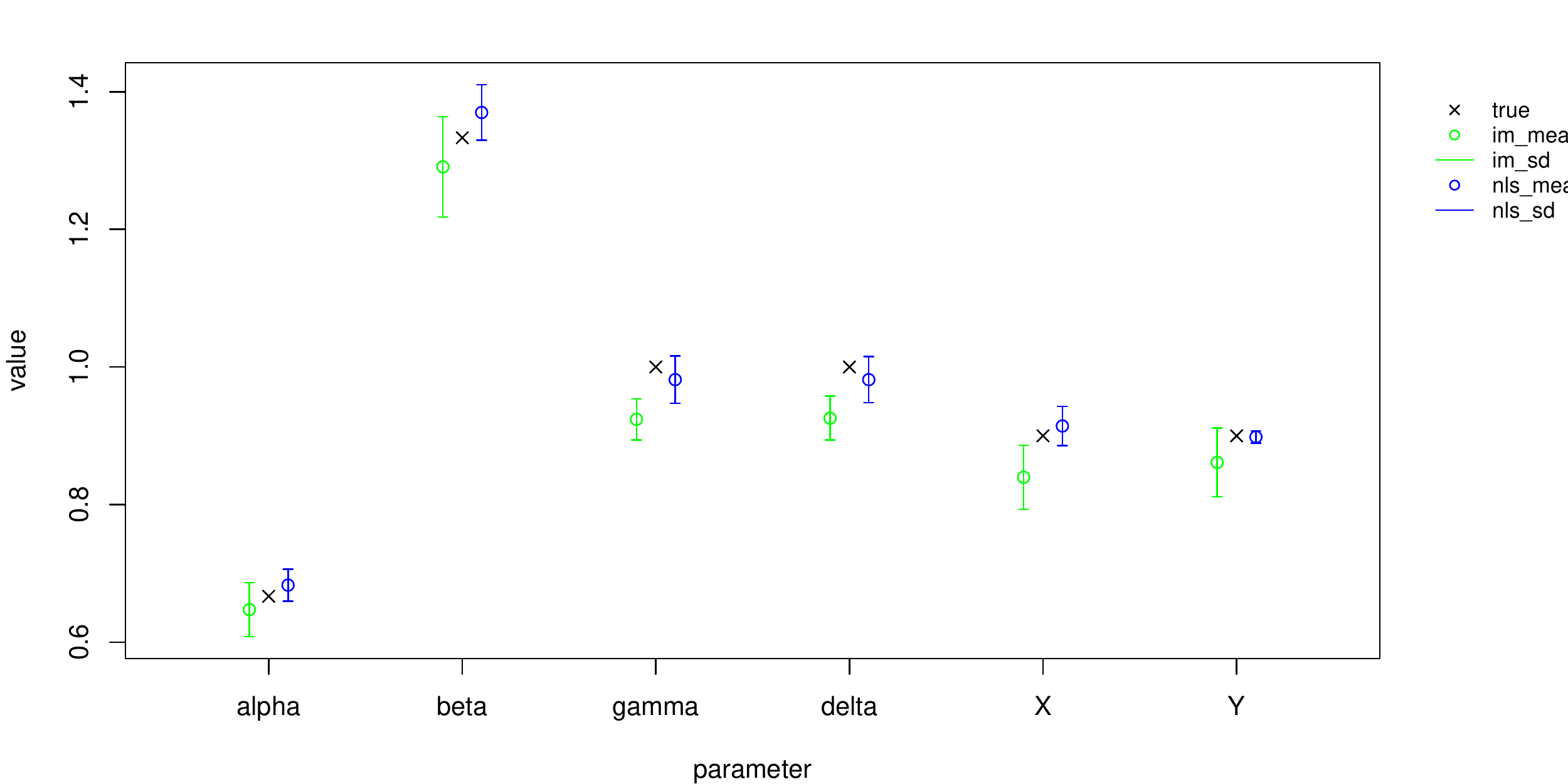}
\caption{\label{fig:lv1} Mean and standard deviation of the parameter estimates from ten Monte-Carlo simulations of the Lotka-Volterra model (\ref{eq:lv}).}
\end{figure}

The returned \code{list.simode} object has its own implementation of plot().
Setting the parameter \code{plot\_mean\_sd=T} the function plots the mean and standard deviation of the fitted curves (not shown here), or parameter estimates as in Figure \ref{fig:lv1}, obtained from the multiple fits.

\subsection{Multiple Subjects}
There are cases where it is reasonable to consider a model where some parameters are assumed to be the same for all experimental subjects while other parameters are specific to an individual subject; see \cite{wang2014estimating} for an example of mixed-effects modeling. While this is possible to do by defining a large ODE model where each individual has his own equations within this model, here we present an easier way to handle such a scenario using the \code{simode} package. We consider the Lotka-Volterra system (two equations) with $N$ individuals. We assume that all individuals share the same system parameter values, whereas each individual has its own initial values. Hence, we would like to use the information from all individuals for estimating the system parameters. We call \code{simode} with the parameter \code{obs} containing a list of length $N$ (where each member of this list is another list containing the observations for this specific subject), and set the parameter \code{obs\_sets} to $N$. We set the control parameter \code{obs\_sets\_fit} in \code{simode.control} to 'separate\_x0' to indicate we want to fit the same parameter values to all subjects but allow different initial conditions. Running \code{simode} returns an object of class \code{list.simode}, where each \code{simode} object contains the estimates for one individual. The parameter estimates in this case will be the same in each of the objects while the initial conditions estimates may be different. In the example below we set $N=5$. Note that one can use the decoupling option here as well by setting \code{decouple\_equations=T}. In this case, the decoupling is designed such that the parameters of each equation will be estimated using the relevant data from all individuals together.

\begin{Schunk}
\begin{Sinput}
R> pars <- c('alpha','beta','gamma','delta')
R> vars <- c('X','Y')
R> eq_X <- 'alpha*X-beta*X*Y'
R> eq_Y <- 'delta*X*Y-gamma*Y'
R> equations <- c(eq_X,eq_Y)
R> names(equations) <- vars
R> theta <- c(2/3,4/3,2,1)
R> names(theta) <- pars
R> n <- 100
R> time <- seq(0,25,length.out=n)
R> N <- 5
R> set.seed(1000)
R> sigma <- 0.05
R> obs <- list()
R> x0_vals <- matrix(NA,N,2)
R> colnames(x0_vals) <- vars
R> for(j in 1:N) {
+    x0 <-rnorm(length(vars),0.9,0.2)
+    x0[x0<0] <- 0
+    x0_vals[j,] <- x0
+    model_out <- solve_ode(equations,theta,x0_vals[j,],time)
+    x_det <- model_out[,vars]
+    obs1 <- list()
+    for(i in 1:length(vars)) {
+      obs1[[i]] <- rnorm(n,x_det[,i],sigma)
+    }
+    names(obs1) <- vars
+    obs[[j]] <- obs1
+  }
R> simode_fits_multi <- simode(
+    equations=equations, pars=c(pars,vars), time=time,
+    obs=obs, obs_sets=N, decouple_equations=T,
+    simode_ctrl=simode.control(obs_sets_fit='separate_x0'))
R> x0_vals


\end{Sinput}
\begin{Soutput}
             X         Y
[1,] 0.8108443 0.6588287
[2,] 0.5635475 0.8533498
[3,] 1.0050157 0.9853895
[4,] 0.7877634 1.1521071
[5,] 1.0325663 0.6240693
\end{Soutput}
\begin{Sinput}
R> summary(simode_fits_multi)
\end{Sinput}
\begin{Soutput}
$im_est
      alpha  beta gamma  delta      X      Y
[1,] 0.6733 1.339 1.971 0.9832 0.9479 0.6576
[2,] 0.6733 1.339 1.971 0.9832 0.6742 0.7109
[3,] 0.6733 1.339 1.971 0.9832 0.8376 1.0550
[4,] 0.6733 1.339 1.971 0.9832 0.8087 1.0640
[5,] 0.6733 1.339 1.971 0.9832 0.8821 0.6594

$nls_est
      alpha  beta gamma delta      X      Y
[1,] 0.6656 1.335 2.007 1.004 0.8107 0.6624
[2,] 0.6656 1.335 2.007 1.004 0.5640 0.8557
[3,] 0.6656 1.335 2.007 1.004 1.0120 0.9946
[4,] 0.6656 1.335 2.007 1.004 0.7907 1.1580
[5,] 0.6656 1.335 2.007 1.004 1.0350 0.6333

$im_loss
[1] 8.829

$nls_loss
[1] 2.377

attr(,"class")
[1] "summary.list.simode"
\end{Soutput}
\end{Schunk}
\subsection{Models with an external input function}

The ODE equations given to \pkg{simode} can include any function of time (e.g., forcing functions) using the reserved symbol 't'. To demonstrate this we expand the predator-prey Lotka-Volterra model from the previous section to include seasonal forcing of the predation rate, using two additional parameters that control the amplitude ($\alpha$) and phase ($\omega$) of the forcing:

\begin{equation}\label{eq:lv_force}
\begin{array}{l}
X^{\prime}(t)=\alpha X(t)-\beta (1+\epsilon\sin (2\pi (t/T+\omega)))X(t)Y(t),
\\
Y^{\prime}(t)=\delta (1+\epsilon\sin (2\pi (t/T+\omega)))X(t)Y(t)-\gamma Y(t).
\end{array}
\end{equation}

The parameter $T$ sets the periodic time scale and is assumed to be known. The following code sets the equations, parameters and the 'true' values for the initial conditions and parameters assuming $T=50$:

\begin{Schunk}
\begin{Sinput}
R> pars <- c('alpha','beta','gamma','delta','epsilon','omega')
R> vars <- c('X','Y')
R> eq_X <- 'alpha*X-beta*(1+epsilon*sin(2*pi*(t/50+omega)))*X*Y'
R> eq_Y <- 'delta*(1+epsilon*sin(2*pi*(t/50+omega)))*X*Y-gamma*Y'
R> equations <- c(eq_X,eq_Y)
R> names(equations) <- vars
R> x0 <- c(0.9,0.9)
R> names(x0) <- vars
R> theta <- c(2/3,4/3,1,1,0.2,0.5)
R> names(theta) <- pars
\end{Sinput}
\end{Schunk}

The following code generates observations from a Gaussian measurement error model:

\begin{Schunk}
\begin{Sinput}
R> n <- 100
R> time <- seq(1,50,length.out=n)
R> model_out <- solve_ode(equations,theta,x0,time)
R> x_det <- model_out[,vars]
R> set.seed(1000)
R> sigma <- 0.1
R> obs <- list()
R> for(i in 1:length(vars)) {
+    obs[[i]] <- x_det[,i] + rnorm(n,0,sigma)
+  }
R> names(obs) <- vars
\end{Sinput}
\end{Schunk}

Attempting to fit the model assuming all parameters are linear fails, as
the parameter $\omega$ is nonlinear in these equations:

\begin{Schunk}
\begin{Sinput}
R> lv_force <- simode(
+    equations=equations, pars=pars, fixed=c(x0), time=time, obs=obs)
\end{Sinput}
\begin{Soutput}
Problem in eq.1 [X] - parameter [omega] should be set as non-linear
Problem in eq.2 [Y] - parameter [omega] should be set as non-linear
\end{Soutput}
\end{Schunk}

Once $\omega$ is defined as nonlinear we encounter a new message:

\begin{Schunk}
\begin{Sinput}
R> nlin_pars <- c('omega')
R> nlin_init <- 0
R> names(nlin_init) <- nlin_pars
R> lv_force <- simode(
+    equations=equations, pars=pars, fixed=c(x0), time=time, 
+    obs=obs, nlin_pars=nlin_pars, start=nlin_init)
\end{Sinput}
\begin{Soutput}
Problem in eq.1 [X] - parameter [beta] or [epsilon] should be 
set as non-linear
Problem in eq.2 [Y] - parameter [delta] or [epsilon] should be 
set as non-linear
\end{Soutput}
\end{Schunk}

\begin{figure}[t!]
\centering
\includegraphics{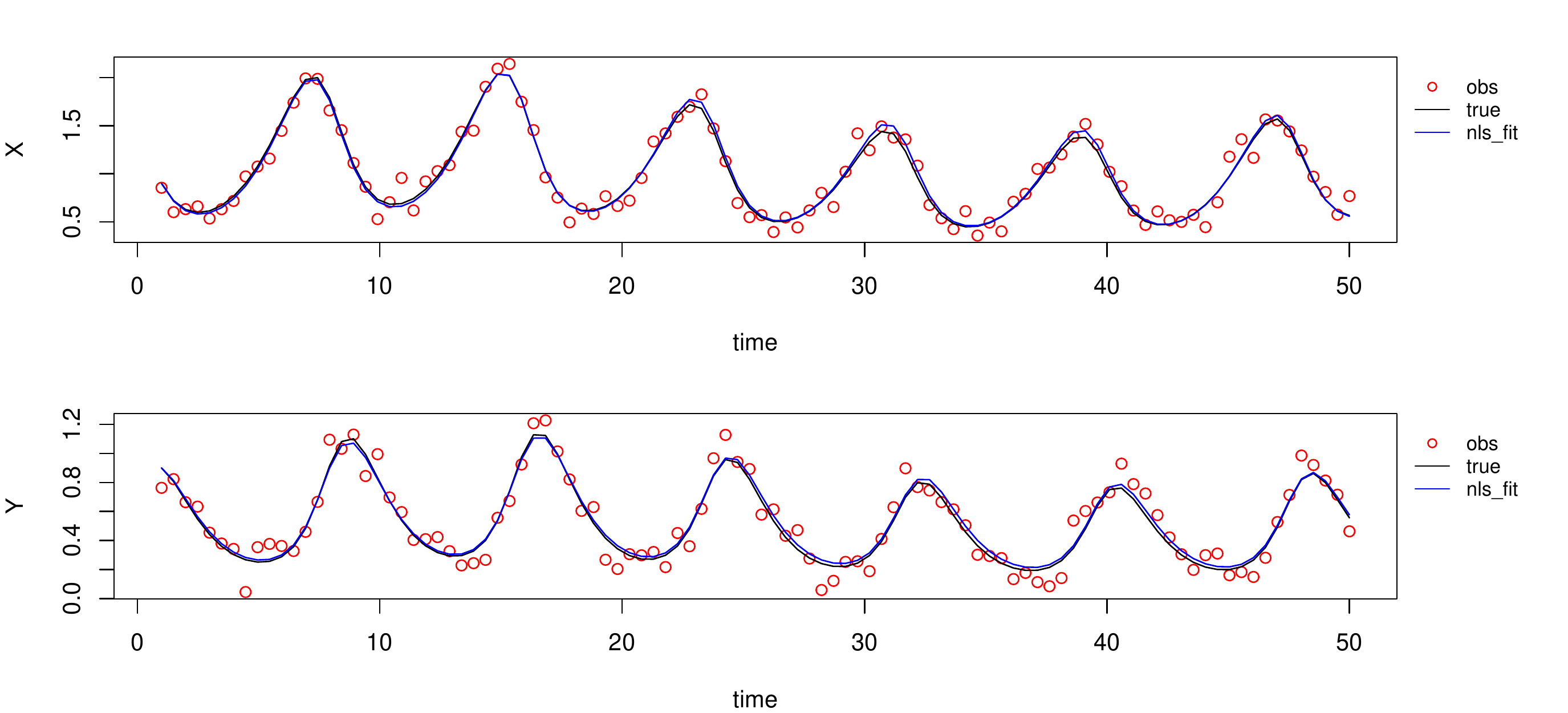}
\caption{\label{fig:lv_force_nm} Model fits according to the Lotka-Volterra equations (\ref{eq:lv_force}) using simplex ('Nelder-Mead') optimization method.}
\end{figure}

Setting $\epsilon$ as nonlinear we can now proceed with the fit.
The package enables the user to modify the optimization methods as is possible in the usual \proglang{R} implementation of \code{optim} function. Here we use the simplex ('Nelder-Mead') method instead of the gradient-based 'BFGS' method (the default), which yields the model fits presented in Figure \ref{fig:lv_force_nm}.

\begin{Schunk}
\begin{Sinput}
R> nlin_pars <- c('epsilon','omega')
R> nlin_init <- c(0.3,0.3)
R> names(nlin_init) <- nlin_pars
R> pars_min <- c(0,0)
R> names(pars_min) <- nlin_pars
R> pars_max <- c(1,1)
R> names(pars_max) <- nlin_pars
R> lv_force1 <- simode(
+    equations=equations, pars=pars, fixed=c(x0), time=time, 
+    obs=obs, nlin_pars=nlin_pars, start=nlin_init, lower=pars_min, 
+    upper=pars_max, simode_ctrl=simode.control(
+			im_optim_method='Nelder-Mead',
+    		nls_optim_method='Nelder-Mead'))
R> summary(lv_force1)$est
\end{Sinput}
\begin{Soutput}
      par       type lower upper start im_est nls_est
1   alpha     linear  -Inf   Inf    NA 0.6288  0.7175
2    beta     linear  -Inf   Inf    NA 1.2210  1.3860
3   gamma     linear  -Inf   Inf    NA 0.9469  0.9269
4   delta     linear  -Inf   Inf    NA 0.9388  0.9138
5 epsilon non-linear     0     1   0.3 0.1709  0.1732
6   omega non-linear     0     1   0.3 0.5174  0.4875
\end{Soutput}
\end{Schunk}

The above example is based on an explicit input function $\sin(t)$. However, one can think of cases where we would like to use a general input, not necessarily for which there is a closed form. In such a case one would incorporate the external input by adding it to the list of observations (the \code{obs} parameter) and referencing it in the equations using the same name:

\begin{Schunk}
\begin{Sinput}
R> pars <- c('alpha','beta','gamma','delta','epsilon')
R> vars <- c('X','Y')
R> eq_X <- 'alpha*X-beta*(1+epsilon*seasonality)*X*Y'
R> eq_Y <- 'delta*(1+epsilon*seasonality)*X*Y-gamma*Y'
R> equations <- c(eq_X,eq_Y)
R> names(equations) <- vars
R> x0 <- c(0.9,0.9)
R> names(x0) <- vars
R> theta <- c(2/3,4/3,1,1,0.2)
R> names(theta) <- pars
R> n <- 100
R> time <- seq(1,50,length.out=n)
R> seasonality <- rep(c(rep(0,10),rep(1,10)),5)
R> xvars <- list(seasonality=seasonality)
R> model_out <- solve_ode(equations,theta,x0,time,xvars)
R> x_det <- model_out[,vars]
R> set.seed(1000)
R> sigma <- 0.1
R> obs <- list()
R> for(i in 1:length(vars)) {
+    obs[[i]] <- x_det[,i] + rnorm(n,0,sigma)
+  }
R> names(obs) <- vars
R> nlin_pars <- c('epsilon')
R> nlin_init <- 0.2
R> names(nlin_init) <- nlin_pars
R> lv_force2 <- simode(
+    equations=equations, pars=pars, fixed=x0, time=time, 
+    obs=c(obs,xvars), nlin_pars=nlin_pars, start=nlin_init)
R> summary(lv_force2)$est
\end{Sinput}
\begin{Soutput}
      par       type start im_est nls_est
1   alpha     linear    NA 0.6824  0.6737
2    beta     linear    NA 1.3090  1.3460
3   gamma     linear    NA 0.9576  0.9912
4   delta     linear    NA 0.9413  0.9906
5 epsilon non-linear   0.2 0.3137  0.2012
\end{Soutput}
\end{Schunk}

\section{Summary} \label{sec:summary}
In this paper we presented the \pkg{simode} \proglang{R} package for conducting statistical inference for ordinary differential equations. The package implements a "two-stage" approach where in the first stage fast estimates of the ODEs parameters are calculated, while the second stage applies nonlinear least squares optimization starting from the estimates obtained. The first stage involves the minimization of an integral criterion function (so called "two-step" approach in the literature) and takes into account separability of parameters and ODEs equations, if such a mathematical feature exists. Several optimization schemes are implemented, depending on the separability characteristics of the ODEs. The package can handle partially observed systems, user defined likelihood functions as well as introducing any function of 't'. Confidence intervals for the ODEs parameters can be calculated as well, hence a full estimation pipeline is implemented.

We demonstrated the usability and flexibility of the package via exploring the parameter estimation of different systems of ODEs. Future abilities are now developed, among them the use of kernel smoothing with automatic bandwidth selection as in \cite{dattnergugushvili18}, and a computational efficient parameter estimation for high dimensional systems of ordinary differential equations.
\section*{Acknowledgments}
This research was supported by the Israeli Science Foundation grant no. 387/15, and by a Grant from the GIF, the German-Israeli Foundation for Scientific Research and Development number I-2390-304.6/2015.

\bibliographystyle{vancouver}
\bibliography{refs}

\begin{thebibliography}{10}

\bibitem{R}
{\proglang{R} Core Team}. \proglang{R}: {A} Language and Environment for
  Statistical Computing.
\newblock Vienna, Austria; 2017.
\newblock Available from: \url{https://www.R-project.org/}.

\bibitem{edelstein2005mathematical}
Edelstein-Keshet L.
\newblock Mathematical models in biology. Classics in Applied Mathematics.
  vol.~46.
\newblock Society for Industrial and Applied Mathematics; 2005.

\bibitem{voit2000computational}
Voit EO.
\newblock Computational analysis of biochemical systems: a practical guide for
  biochemists and molecular biologists.
\newblock Cambridge University Press; 2000.

\bibitem{anderson1992infectious}
Anderson RM, May RM, Anderson B.
\newblock Infectious diseases of humans: dynamics and control. vol.~28.
\newblock Wiley Online Library; 1992.

\bibitem{hooker2015collocinfer}
Hooker G, Ramsay JO, Xiao L, et~al.
\newblock CollocInfer: Collocation Inference in Differential Equation Models.
\newblock Journal of Statistical Software. 2015;.

\bibitem{ramsay2007parameter}
Ramsay JO, Hooker G, Campbell D, Cao J.
\newblock Parameter estimation for differential equations: a generalized
  smoothing approach.
\newblock Journal of the Royal Statistical Society: Series B (Statistical
  Methodology). 2007;69(5):741--796.

\bibitem{raue2015data2dynamics}
Raue A, Steiert B, Schelker M, Kreutz C, Maiwald T, Hass H, et~al.
\newblock Data2Dynamics: a modeling environment tailored to parameter
  estimation in dynamical systems.
\newblock Bioinformatics. 2015;31(21):3558--3560.

\bibitem{mikkelsen2017learning}
Mikkelsen FV, Hansen NR.
\newblock Learning Large Scale Ordinary Differential Equation Systems.
\newblock arXiv preprint arXiv:171009308. 2017;.

\bibitem{wood2010statistical}
Wood SN.
\newblock Statistical inference for noisy nonlinear ecological dynamic systems.
\newblock Nature. 2010;466(7310):1102.

\bibitem{wilkinson2011package}
Wilkinson D.
\newblock Package 'smfsb'. 2011;.

\bibitem{JSSv069i12}
King A, Nguyen D, Ionides E.
\newblock Statistical Inference for Partially Observed Markov Processes via the
  R Package pomp.
\newblock Journal of Statistical Software, Articles. 2016;69(12):1--43.
\newblock Available from: \url{https://www.jstatsoft.org/v069/i12}.

\bibitem{wu2008dediscover}
Wu H, Miao O, Warnes GR, Wu C, LeBlanc A, Dykes C, et~al.
\newblock Dediscover: a computation and simulation tool for hiv viral fitness
  research.
\newblock In: BioMedical Engineering and Informatics, 2008. BMEI 2008.
  International Conference on. vol.~1. IEEE; 2008. p. 687--694.

\bibitem{dattner2015}
Dattner I, Klaassen CAJ.
\newblock Optimal rate of direct estimators in systems of ordinary differential
  equations linear in functions of the parameters.
\newblock Electron J Statist. 2015;9(2):1939--1973.
\newblock Available from: \url{http://dx.doi.org/10.1214/15-EJS1053}.

\bibitem{dattner2015model}
Dattner I.
\newblock A model-based initial guess for estimating parameters in systems of
  ordinary differential equations.
\newblock Biometrics. 2015;71(4):1176--1184.

\bibitem{dattner2017modelling}
Dattner I, Miller E, Petrenko M, Kadouri DE, Jurkevitch E, Huppert A.
\newblock Modelling and parameter inference of predator--prey dynamics in
  heterogeneous environments using the direct integral approach.
\newblock Journal of The Royal Society Interface. 2017;14(126):20160525.

\bibitem{yaarietal18}
Yaari R, Dattner I, Huppert A.
\newblock A two-stage approach for estimating the parameters of an age-group
  epidemic model from incidence data.
\newblock Statistical Methods in Medical Research. 2018;27(7):1999--2014.
\newblock PMID: 29260611.
\newblock Available from: \url{https://doi.org/10.1177/0962280217746443}.

\bibitem{dattnergugushvili18}
Dattner I, Gugushvili S.
\newblock Application of one-step method to parameter estimation in ODE models.
\newblock Statistica Neerlandica. 2018;72(2):126--156.

\bibitem{vujavcic2015time}
Vuja{\v{c}}i{\'c} I, Dattner I, Gonz{\'a}lez J, Wit E.
\newblock Time-course window estimator for ordinary differential equations
  linear in the parameters.
\newblock Statistics and Computing. 2015;25(6):1057--1070.

\bibitem{vujavcic2018consistency}
Vuja{\v{c}}i{\'c} I, Dattner I.
\newblock Consistency of direct integral estimator for partially observed
  systems of ordinary differential equations.
\newblock Statistics \& Probability Letters. 2018;132:40--45.

\bibitem{voit2004decoupling}
Voit EO, Almeida J.
\newblock Decoupling dynamical systems for pathway identification from
  metabolic profiles.
\newblock Bioinformatics. 2004;20(11):1670--1681.

\bibitem{Bellman197126}
Bellman R, Roth RS.
\newblock The use of splines with unknown end points in the identification of
  systems.
\newblock Journal of Mathematical Analysis and Applications. 1971;34(1):26--33.

\bibitem{varah1982spline}
Varah J.
\newblock A spline least squares method for numerical parameter estimation in
  differential equations.
\newblock SIAM Journal on Scientific and Statistical Computing.
  1982;3(1):28--46.

\bibitem{brunel2008parameter}
Brunel NJB.
\newblock Parameter estimation of ODE's via nonparametric estimators.
\newblock Electronic Journal of Statistics. 2008;2:1242--1267.

\bibitem{liang2008parameter}
Liang H, Wu H.
\newblock Parameter estimation for differential equation models using a
  framework of measurement error in regression models.
\newblock Journal of the American Statistical Association.
  2008;103(484):1570–--1583.

\bibitem{fang2011two}
Fang Y, Wu H, Zhu LX.
\newblock A two-stage estimation method for random coefficient differential
  equation models with application to longitudinal HIV dynamic data.
\newblock Statistica Sinica. 2011;21(3):1145.

\bibitem{gugushvili2012sqrt}
Gugushvili S, Klaassen CAJ.
\newblock $\sqrt{n}$-consistent parameter estimation for systems of ordinary
  differential equations: bypassing numerical integration via smoothing.
\newblock Bernoulli. 2012;18:1061--1098.

\bibitem{ramsay2017dynamic}
Ramsay J, Hooker G.
\newblock Dynamic Data Analysis: Modeling Data with Differential Equations.
\newblock Springer; 2017.

\bibitem{golub2003separable}
Golub G, Pereyra V.
\newblock Separable nonlinear least squares: the variable projection method and
  its applications.
\newblock Inverse problems. 2003;19(2):R1.

\bibitem{chou2009recent}
Chou IC, Voit EO.
\newblock Recent developments in parameter estimation and structure
  identification of biochemical and genomic systems.
\newblock Mathematical biosciences. 2009;219(2):57.

\bibitem{deSolve}
Soetaert K, Petzoldt T, Setzer RW.
\newblock Solving Differential Equations in R: Package deSolve.
\newblock Journal of Statistical Software. 2010;33(9):1--25.
\newblock Available from: \url{http://www.jstatsoft.org/v33/i09}.

\bibitem{wang2014estimating}
Wang L, Cao J, Ramsay J, Burger D, Laporte C, Rockstroh J.
\newblock Estimating mixed-effects differential equation models.
\newblock Statistics and Computing. 2014;24(1):111--121.

\end{thebibliography}


\end{document}